\documentclass[a4paper,11pt]{article}

\pdfoutput=1 

\usepackage{jheppub} 
\usepackage{amsmath,amssymb,mathtools}
\usepackage[T1]{fontenc}
\usepackage{graphicx}
\usepackage{epsfig}
\usepackage{nicefrac}
\usepackage{tabularx}
\usepackage{bbm,bm}
\usepackage{enumitem}

\newcolumntype{L}[1]{>{\raggedright\arraybackslash}p{#1}}
\newcolumntype{C}[1]{>{\centering\arraybackslash}p{#1}}
\newcolumntype{R}[1]{>{\raggedleft\arraybackslash}p{#1}}

\renewcommand{\arraystretch}{1.2}

\title{\boldmath The impact of transverse Slavnov-Taylor identities on dynamical chiral symmetry breaking}

\author[a]{Luis Albino,}
\author[b]{Adnan Bashir,}
\author[c1]{Bruno El-Bennich,\note{\, Corresponding author}}
\author[d]{Eduardo Rojas,}
\author[ae]{Fernando E. Serna}
\author[c]{and Roberto Correa da Silveira}

\affiliation[a]{Instituto de F\'isica Te\'orica, Universidade Estadual Paulista, Rua Dr. Bento Teobaldo Ferraz, 271, 01140-070 S\~ao Paulo, SP, Brazil}
\affiliation[b]{Instituto de F\'isica y Matem\'aticas, Universidad Michoacana de San Nicol\'as de Hidalgo, Morelia, Michoac\'an 58040, Mexico}
\affiliation[c]{Laborat\'orio de F\'isica Te\'orica e Computacional, Universidade Cidade de S\~ao Paulo, Rua Galv\~ao Bueno 868, 01506-000 S\~ao Paulo, SP, Brazil}
\affiliation[d]{Departamento de F\'isica, Universidad de Nari\~no, A.A. 1175, San Juan de Pasto, Colombia}
\affiliation[e]{Departamento de F\'isica, Universidad de Sucre, Carrera 28 No. 5-267, Barrio Puerta Roja, Sincelejo, Colombia}

\emailAdd{bruno.elbennich@cruzeirodosul.edu.br}

\abstract{We extend earlier studies of transverse Ward-Fradkin-Green-Takahashi identities in QED, their usefulness to constrain the transverse fermion-boson vertex and
               their importance for multiplicative renormalizability, to the equivalent gauge identities in QCD. To this end, we consider transverse Slavnov-Taylor identities that
               constrain the transverse quark-gluon vertex and derive its eight associated scalar form factors. The complete vertex can be expressed in terms of the quark's
               mass and wave-renormalization functions, the ghost-dressing function, the quark-ghost scattering amplitude and a set of eight form factors. The latter parametrize
               the hitherto unknown nonlocal tensor structure in the transverse Slavnov-Taylor identity which arises from the Fourier transform of a four-point function involving 
               a Wilson line in coordinate space. We determine the functional form of these eight form factors with the constraints provided by the Bashir-Bermudez vertex and 
               study the effects of this novel vertex on the quark in the Dyson-Schwinger equation using lattice QCD input for the gluon and ghost propagators.
               We observe significant dynamical chiral symmetry breaking and a mass gap that leads to a constituent mass of the order of 500~MeV for the light quarks.
               The flavor dependence of the mass and wave-renormalization functions as well as their analytic behavior on the complex momentum plane is studied and as an
               application we calculate the quark condensate and the pion's weak decay constant in the chiral limit. Both are in very good agreement with their reference values. }

\begin{document}
\maketitle
\flushbottom

\section{Introduction}
\label{sec:intro}

Whilst the Brout-Englert-Higgs (BEH) mechanism  has been established as the essential explicit source of elementary particles' masses, the same cannot be said of
Nature's composite building blocks, namely the atoms and their nuclei. Even the lightest Nambu-Goldstone mode of Quantum Chromodynamics (QCD), the pion, is
more than an order of magnitude heavier than the sum of the current masses of its constituents provided  by the BEH mechanism. QCD is not a conformal
theory and this mass relation is only sensible at a certain energy scale. Commonly, the light quark's current masses are quoted at 2~GeV in the $\overline {\rm MS}$ 
scheme~\cite{Aoki:2019cca}, at which the sum of two $u$ and one $d$ current-quark current masses amounts to merely 1\% of the proton mass.

The overwhelming contribution to the light hadron's masses does not stem from the aforementioned mechanism and a recent lattice-QCD simulation~\cite{Yang:2018nqn}
concludes that the kinetic quark energy and the gluon field are together responsible for 68\% of the proton mass, whereas the trace anomaly contributes 23\%.
Thus, only 9\% of the proton's mass is due to the scalar condensate, which exhibits the major current-quark mass dependence. This recent analysis of the proton's
``mass budget'' has come to strengthen long-standing observations made by groups who apply functional continuum methods to QCD. In particular,  starting from light
current quarks with $m_{u,d} \approx$~3--4~MeV at a reasonable perturbative scale, solving the Dyson-Schwinger equation (DSE) for the quark propagator in
QCD~\cite{Roberts:1994dr,Bashir:2012fs} and making use of the three-body Faddeev equation yields the proton's mass and that of the Roper, the nucleon's parity
partners and the $\Delta$ baryons in a  consistent symmetry-preserving truncation~\cite{Cloet:2008re,Eichmann:2009qa,Aznauryan:2012ba,Segovia:2015hra,
Eichmann:2016hgl,Eichmann:2016yit,Chen:2017pse,Sanchis-Alepuz:2017jjd,Chen:2018nsg,Bednar:2018htv}. These results not only preserve the correct mass
ordering but are also remarkably accurate  within the uncertainties of the meson cloud effect. In these approaches one can also compute the nucleon's $\sigma$
term, which is a measure of its current-quark mass dependence~\cite{Flambaum:2005kc}, and it turns out that $\sigma_N \simeq$~50--60~MeV. The evidence
is conclusive that the vast majority of the nucleon's mass is due to \emph{dynamical chiral symmetry breaking\/} (DCSB), which is in contrast to explicit chiral
symmetry breaking whose origin lies solely in  the BEH mechanism and is described by the quarks' mass terms in the QCD Lagrangian.

The Faddeev bound-state calculations are limited to the leading approximation of the gap equation and three-body interaction kernel, with the exception of a certain
degree of beyond-ladder truncations effects modeled in the diquark amplitudes and propagators. In this simplification, referred to as \emph{rainbow-ladder} truncation,
the quark-gluon vertex is simply described by its bare form defined in the Lagrangian coupling. This is a drastic simplification, as the one-loop dressed
vertex already leads to non-vanishing coefficients for all 12 independent tensor structures~\cite{Ball:1980ay,Davydychev:2000rt}. In fact, DCSB is manifest not only
in the quark propagator but also in the quark-gluon vertex, as 6 of the 12 structures are only generated dynamically and their feed-back into the gap equation enhances
the generation of quark masses. If one solves the gap equation with merely the bare vertex, a realistic strong coupling and a gluon propagator obtained with the appropriate
DSE or lattice-QCD simulations~\cite{Fischer:2008uz,Alkofer:2008jy,Dudal:2008sp,Aguilar:2004sw,Aguilar:2008xm,Aguilar:2012rz,Cucchieri:2007md,Cucchieri:2007rg,
Oliveira:2008uf,Pennington:2011xs,Oliveira:2012eh,Bogolubsky:2009dc,Ayala:2012pb,Strauss:2012dg,Cyrol:2016tym,Boucaud:2018xup,Mintz:2018hhx,Duarte:2016iko,
Dudal:2018cli,Aguilar:2019uob,Huber:2015ria,Huber:2020keu,Fischer:2020xnb,Falcao:2020vyr}, the resulting DCSB is too small and practically irrelevant to realistic
constituent-quark and hadron masses. Therefore, in using a model for the gluon dressing in the rainbow-ladder truncation, a scale parameter is adjusted to the
experimental pion and kaon masses.

This procedure has proven to be very successful for ground and first excited states of mesons and quarkonia in the pseudoscalar and vector channels. However, higher
excited and exotic states, as well as the scalar and axialvector $\bar qq$ states, are not well described in this leading truncation~\cite{Cloet:2008re,Eichmann:2009qa,
Aznauryan:2012ba,Segovia:2015hra,Eichmann:2016hgl,Eichmann:2016yit,Chen:2017pse,Sanchis-Alepuz:2017jjd,Chen:2018nsg,Bednar:2018htv,Maris:1997tm,
Maris:1997hd,Maris:1999nt,Alkofer:2002bp,Qin:2011dd,Chang:2011ei,Rojas:2014aka,Raya:2015gva,El-Bennich:2016qmb,Mojica:2017tvh,Serna:2017nlr,Bedolla:2015mpa,
Bedolla:2016yxq,Raya:2017ggu,Fischer:2014cfa,Hilger:2017jti,Gunkel:2019xnh}. A particularly problematic case is the highly asymmetric momentum distribution in
flavored mesons, such as the $D$ and $B$ mesons, as the quark-gluon vertex dressing has a different impact on a light quark than on charm and bottom quarks.
More precisely, while a bare vertex can reliably be used for the heavy-quark interaction in the Bethe-Salpeter equation (BSE), this is not the case for light
quarks~\cite{ElBennich:2008qa,ElBennich:2009vx,El-Bennich:2017brb,ElBennich:2012tp,AtifSultan:2018end,Serna:2018dwk,Serna:2020txe}. Effects of a dressed 
quark-gluon vertex in heavy-light mesons have also been addressed in Refs.~\cite{Gomez-Rocha:2014vsa,Gomez-Rocha:2015qga,Gomez-Rocha:2016cji}.

There are not only phenomenological motivations for improving our knowledge about the tensor structure and analytic behavior of the quark-gluon vertex in the
nonperturbative domain. On pure field-theoretical grounds the behavior of the fermion-boson vertex is of great interest, as it \emph{i}) is a crucial object to explain
DCSB in QED (with an artificially scaled-up coupling) and in QCD, and \emph{b}) plays an eminent role in its contribution to the infrared behavior of Green functions
as well as to quark and gluon fragmentation functions, and therefore to the elucidation of the confinement mechanism. Starting from the perturbative limit,
truncation models for the vertex are commonly constructed that retain the essential features of the underlying theory, which are known to be respected at
every order of a perturbative formulation by construction. It is a natural requirement to achieve a satisfactory determination of relevant physical observables.
For small values of the coupling constant, perturbation theory is the paradigmatic example of such a truncation scheme, yet inadequate to calculate hadronic
bound states. Much progress towards a more detailed understanding of the fermion-boson vertex in QED and QCD  has been made in the past years,
either by direct calculation of the DSE for the vertex~\cite{Fischer:2003rp,Alkofer:2008tt,Hopfer:2013np,Williams:2014iea,Williams:2015cvx},
invoking gauge identities and multiplicative renormalizability~\cite{Curtis:1990zs,Bashir:1994az,Bashir:1995qr,Bashir:2004yt,
Bashir:2004mu,Kizilersu:2009kg,Bashir:2011vg,Bashir:2011dp,Aslam:2015nia,Fernandez-Rangel:2016zac,Bermudez:2017bpx,Albino:2018ncl,Rojas:2013tza,Rojas:2014tya,
Qin:2013mta,Binosi:2016wcx,Aguilar:2010cn,Aguilar:2014lha,Aguilar:2016lbe,Aguilar:2018epe,Aguilar:2018csq,Pennington:2016vxv,Oliveira:2018ukh,Oliveira:2020yac},
in perturbative approaches~\cite{Bashir:1997qt,Bashir:1999bd,Bashir:2000rv,Pelaez:2015tba} or by direct numerical sampling of QCD on the lattice~\cite{Skullerud:2002ge,
Skullerud:2003qu,Kizilersu:2006et,Oliveira:2016muq,Sternbeck:2017ntv,Kizilersu:2021jen}.

In this work we extend earlier studies~\cite{Albino:2018ncl} which are based on gauge covariance and combine the constraints of the well known
Ward-Fradkin-Green-Takahashi (WFGTI) identity~\cite{Ward:1950xp,Fradkin:1955jr,Green:1953te,Takahashi:1957xn} \emph{and two additional\/} transverse
Takahashi identities (TTI)~\cite{Takahashi:1985yz,Kondo:1996xn,He:2000we,He:2006my,He:2007zza}. The WFGTI has long been known as an expression of gauge
symmetry and current conservation and allows for an expansion of the photon-fermion vertex in terms of a well-constrained ``\emph{longitudinal\,}''
part\footnote{\,This denomination of the non-transverse vertex is a misnomer, as a purely transverse gluon propagator in Landau gauge projects
out any longitudinal contributions of  the vertex in the DSE kernel.}~\cite{Ball:1980ay} and undetermined transverse components. The symmetry which leads to the
transverse identities is the Lorentz transformation acting on the usual infinitesimal gauge transformation. Indeed, while the WFGTI relates the divergence of the
fermion-photon vertex to the inverse fermion propagator, the TTI expresses the curl of this vertex. Though the TTI were verified to one-loop
order~\cite{Pennington:2005mw,He:2006ce}, the complexity of these additional structures made these identities for the longest  time not amicable to a straightforward
determination of the transverse part of the photon-fermion vertex. Moreover, the TTI couple the vector and axialvector vertices, though, as demonstrated in
Ref.~\cite{Qin:2013mta}, the uncoupling can be achieved by judicious tensor projections and leaves one with two identities that involve only the vector
vertex.\footnote{\,Similar projections lead to transverse identities that only involve the axialvector vertex.} Nonetheless, besides the inverse fermion  propagator
and the vector vertex, these uncoupled TTI still involve a nontrivial tensor structure that arises from the Fourier transform of a four-point like function in
coordinate-space with a necessary Wilson line.

In Ref.~\cite{Albino:2018ncl} we demonstrated that this tensor structure can be parametrized and \emph{is constrained by multiplicative renormalizability\/}.
In that capacity, the TTI are intimately connected to another consequence of local gauge covariance, namely the Landau-Khalatnikov-Fradkin transformations
(LKFT)~\cite{Landau:1955zz,Fradkin:1955jr} which describe the response of the Green functions to an arbitrary gauge transformation and express multiplicative
renormalizability  of the massless fermion propagator in 4 space-time dimensions. This implies that not any functional form of the tensor structures in the transverse
vertex is possible and we showed that, for a given form of the transverse vertex that satisfies the TTI,  the critical QED coupling above which chiral symmetry is
dynamically broken is gauge invariant.

We here build upon these  results and  explore them in the context of QCD. As it is well known, color-gauge invariance in  QCD is preserved by the Slavnov-Taylor
identity (STI) for the quark-gluon vertex~\cite{Slavnov:1972fg,Taylor:1971ff} which \emph{also\/} leaves the transverse vertex  undetermined. The TTI were generalized
to two transverse STIs (TSTI)~\cite{He:2009sj} from which the vector vertex can be extracted that involves, as the usual STI, the  inverse quark propagators, the 
ghost-dressing function and  the quark-ghost scattering amplitude, but furthermore  a nonlocal four-point like function which is  a consequence of gauge invariance.
As in QED, this latter term can be parametrized most generally by four tensor structures and corresponding form factors.  Similarly, the quark-ghost scattering kernel
can be described by four matrix-valued amplitudes which can be  computed within a nonperturbative {\em dressed}-propagator
approach~\cite{Rojas:2013tza,Aguilar:2010cn,Aguilar:2016lbe,Aguilar:2018epe,Aguilar:2018csq}. Note that, for the very first time, we now also have generalized
LKFT (GLKFT) for QCD~\cite{Aslam:2015nia,DeMeerleer:2018txc,DeMeerleer:2019kmh}, though our understanding of these is in its infancy and their complexity
still prevents us from imposing tangible constraints on the quark-gluon vertex.

In analogy with the approach taken in Refs.~\cite{Qin:2013mta,Albino:2018ncl}, we divide the quark-gluon vertex into longitudinal and transverse tensor structures using the 
tensor basis of Ref.~\cite{Kizilersu:1995iz}. Solving a system of coupled equations we obtain the expressions of the longitudinal form factors  $\lambda_i (k,p), i=1,...,4$,  from 
the usual STI as in Refs.~\cite{Rojas:2013tza,Aguilar:2010cn,Aguilar:2016lbe,Aguilar:2018epe,Aguilar:2018csq}, and additionally derive the functional form of the transverse
form factors $\tau_j(k,p), j =1,...,8$, where $k$ and $p$ are the outgoing and incoming quark momenta, respectively. Six of the twelve vertex form factors depend on the quark-ghost
interaction kernel and on the ghost-dressing function. In addition, the transverse form factors are characterized by a functional dependence on eight scalar functions,
$Y_i(k,p)$, which parametrize the aforementioned nonlocal tensor structure in the uncoupled TSTI for the vector vertex.  As we are currently in no condition to calculate
the corresponding four-point function in momentum space, we rely on the well established  Bashir-Bermudez  ansatz~\cite{Bashir:2011dp} for the fermion-gauge boson
vertex that preserves multiplicative renormalizability and is constrained by  gauge covariance and perturbative QCD in a given kinematic limit. Thus, we trade our
ignorance of the $Y_i(k,p)$ functions for eight parameters $a_i$ employed in this ansatz and solve the quark DSE with the complete vertex structure for a large
representative sample of $a_i$-sets. The latter are constrained by multiplicative renormalizability in the range, $-2 \leq a_i\leq +2$, and we employ gluon
and ghost  propagators  from lattice QCD.

We find that only a very limited combination leads to a mass function $M(p)$ that exhibits the DCSB and functional behavior observed in phenomenological models,
and the same holds for the wave renormalization function $Z(p)$. In many cases, uninteresting solutions are found, i.e. $M(0) \lesssim 100$~MeV, or the iteration
process to solve the integral equations converges poorly or not at all. We here present the first solution of the quark's DSE that leads to significant DCSB with
$M(0) \approx  500$~MeV, employing $\alpha_s (\mu) \simeq 0.3$, a gluon-dressing function, $\Delta (q^2, \mu$), and a ghost-dressing function, $G(q^2,\mu^2)$,
renormalized  at $\mu = 4.3$~GeV in  agreement with lattice QCD. No additional strength via a form factor or other modifications in the DSE kernel are introduced and
even in the chiral limit the DCSB is still considerable.

This paper is organized as follows: in Section \ref{sec2} we review the DSE for a quark in QCD and motivate our renormalization procedure, after which
we go into the details of the construction of the fully dressed quark-gluon vertex and its most general tensor structure. We then introduce the longitudinal
and transverse STIs that constrain the quark-gluon vertex, discuss their content, in particular the quark-ghost scattering kernel,  and show how their manipulations
with appropriate tensor projections leads to the analytic form of the 12 associated form factors. Since one nonlocal tensor structure in the transverse STIs is
hitherto unknown, at least nonperturbatively, we introduce its parametrization and constrain it with the Bashir-Bermudez ansatz in Ref.~\cite{Bashir:2011dp}
and multiplicative renormalizability. In Section~\ref{ghostgluonsec} we turn our attention to the gauge sector and list the  parametrizations of the gluon
and ghost propagators which are fitted to the data from three lattice QCD groups. In Section~\ref{sec4} we present our main results which consist of
the mass and wave-renormalization functions and the leading quark-ghost form factor in case of omitting the transverse vertex  and using the full vertex
in the DSE. Strong coupling variations and quark-flavor dependence of the DSE solutions are also studied. We wrap up that section with the solutions
of the DSE on the complex plane for light quarks, followed by some applications of our results, namely the calculation of the pion's weak decay constant and
the quark condensate in the chiral limit, in Section~\ref{sec5}. Finally, in Section~\ref{sec6} we comment on our results and propose future steps that could
lead to a parameter-free determination of all transverse vertex functions.


\section{The gap equation and gauge-symmetry constraints on the quark-gluon vertex}
\label{sec2}


\subsection{Dyson-Schwinger equation for a quark \label{DSE}}

The most prominent occurrence of the quark-gluon vertex is in the DSE, which is nothing else than the relativistic equation of motion of the quark in QCD formulated
in a nonperturbative manner. In essence, the DSE describes the nonperturbative gluon dressing of the current quark by a self-energy term in its propagator. For a given
flavor the DSE of the inverse quark propagator is,\footnote{\,We work in Euclidean space in which $\left\{ \gamma_{\mu},\gamma_{\nu} \right\} =
2 \delta_{\mu\nu}$, where $\delta_{\mu\nu}$ is the Euclidean metric and the Dirac matrices are hermitian: $\gamma^{\dagger}_{\mu}=\gamma_{\mu}$. Moreover,
$\gamma_{5}= \gamma_5^\dagger = \gamma_{4} \gamma_{1} \gamma_{2} \gamma_{3}$,
 with  $\mathrm{Tr}\left[ \gamma_{5} \gamma_{\mu} \gamma_{\nu} \gamma_{\alpha} \gamma_{\beta} \right] = -4 \epsilon_{\mu \nu \alpha \beta}$,
$\sigma_{\mu \, \nu} = \frac{i}{2} \left[ \gamma_{\mu},\gamma_{\nu} \right ]$ and a space-like vector $p_\mu$ is characterized by $p^2 >0$. }
\begin{align}
  S^{-1}(p) & = \,  Z_2 \, \left ( i\, \gamma\cdot p + m^{\mathrm{bm}}\right  )   + \Sigma (p^2) \nonumber \\
                 & = \, Z_2 \,  i\, \gamma\cdot p + Z_4 \, m  + Z_1\, g^2\!  \int^\Lambda\!\!  \frac{d^4k}{(2\pi)^4}\  \Delta^{ab}_{\mu\nu} (q)\, \gamma_\mu t^a\,  S(k)\,\Gamma_\nu^b (k,p) \ ,
 \label{DSEquark}
\end{align}
where $m^{\mathrm{bm}}$ is the bare quark mass, $m$ is the renormalized or current quark mass and $Z_1(\mu,\Lambda)$ and $Z_2(\mu,\Lambda)$
are the vertex and wave-function  renormalization constants, respectively. The first two terms in Eq.~\eqref{DSEquark} are the inverse free quark propagator and the
integral expresses the quark's self-energy $\Sigma (p^2)$. In this integral, $D_{\mu\nu}(q)$ is the dressed-gluon propagator in Landau gauge with momentum $q=k-p$,
\begin{equation}
    \Delta^{ab}_{\mu\nu} (q ) =  \delta^{a b} \left(\delta_{\mu \nu}-\frac{q_{\mu} q_{\nu}}{q^{2}}\right ) \! \Delta (q^2) \ ,
 \label{gluonprop}
\end{equation}
where $\Delta(q^2)$ is the nonperturbative dressing function, $\Delta (q^2 ) \to 1/q^2$ for large $q^2$,  $\Gamma^a_\mu (k,p) = \Gamma_\mu (k,p)\, t^a $,
is the dressed quark-gluon vertex and $t^a = \lambda^a/2$  are the SU(3) group generators with $\lambda^a$  in the fundamental
representation; $a,b$ generally represent color indices.

The most general Poincar\'e-covariant form of the solutions to Eq.~\eqref{DSEquark} is written in terms of covariant scalar and vector amplitudes:
\begin{eqnarray}
\label{DEsol}
   S (p) \, & =  & -i \gamma \cdot p \, \sigma_{\rm v} ( p^2 ) + \sigma_{\rm s} ( p^2 )    \nonumber \\
               & = &  \frac{1}{i \gamma \cdot p \,A (p^2)   + B ( p^2 ) } = \,  \frac{Z (p^2 )}{ i \gamma \cdot p + M ( p^2 )} \   .
\end{eqnarray}
In the integral, $\Lambda$ is an ultraviolet cut-off, $\mu$ is the renormalization scale and one typically chooses $\Lambda \gg \mu$ in DSE studies of QCD.
These scales are implicit in our notation, namely $A(p^{2}) \equiv A(p^{2},\mu^2,\Lambda^2)$ and $B(p^{2}) \equiv B(p^{2},\mu^2,\Lambda^2)$, as is a flavor index
$f$ for these quantities as well as for all renormalization constants. The flavor-dependent nonperturbative mass and wave renormalization functions are,
\begin{equation}
      M (p^2) = B(p^{2},\mu^2,\Lambda^2)/A(p^{2},\mu^2,\Lambda^2) \ ,
\end{equation}
and,
\begin{equation}
     Z(p^2,\Lambda^2,\mu^2) = 1/A(p^2,\Lambda^2,\mu^2) \ ,
\end{equation}
respectively. The scale $\mu$ is commonly chosen such that the dressed functions match the ones in perturbation theory, i.e. $Z(\mu^2) =1$ and $M(\mu^2 ) =
m =  Z_2(\mu,\Lambda)/ Z_4(\mu,\Lambda) \,m^{\mathrm{bm}} (\Lambda)$, where $Z_4(\mu,\Lambda)$ is the renormalization constant associated with the Lagrangian's
mass term.

The object of our interest is the fully dressed vertex, $\Gamma_\mu (k,p)$, which satisfies its own DSE. As already mentioned, we do not intend to solve
the DSE of the vertex in a given truncation scheme but rather make use of three STIs presented in Sections~\ref{generaltensor} and \ref{TSTIs}. This allows us
to derive the scalar form factors associated with the tensor structure of the vertex. Before we engage on this path, a note on the renormalization method is in order.
After taking the color trace and with $C_F = 4/3$ in the fundamental representation, the unrenormalized  DSE for a given quark reads:
\begin{equation}
   S^{-1}_0(p) = i \gamma\cdot p +  m^{\mathrm{bm}}  + g_0^2\, C_F\! \int\! \frac{d^4k}{(2\pi)^4}\  \Delta^0_{\mu\nu}(q) \gamma_\mu\, S_0(k)\,\Gamma_\nu^0 (k,p) \ .
\end{equation}
Relating bare propagators, the coupling and vertex to their renormalized expressions via the following procedure,
\begin{spreadlines}{10pt}
\begin{align}
    S (p,\mu^2 ) & = \,  Z_2^{-1}(\mu^2 )  S_0 (p) \ , \\
    \Delta (q^2, \mu^2 ) & = \,  Z_{A}^{-1} (\mu^2 ) \Delta_0 (q^{2} )  \ ,    \\
    \Gamma_\mu (k, p, \mu^2 ) &  =  \,  Z_1(\mu^2 )\, \Gamma_\mu^0 (k, p) \ , \\
     g (\mu^2 ) & =  \, Z_g^{-1} (\mu^2 )\, g_0 \ ,
\end{align}
and considering the relation $Z_g^2 = Z_1^2/(Z_2^2Z_A)$ for the strong coupling~\cite{Itzykson:1980rh} we obtain:
\begin{equation}
  Z_2^{-1} S^{-1}(p) = i \gamma\cdot p + m^{\mathrm{bm}}
                              +  \frac{C_F g^2 Z_1^2}{Z_2^2 Z_A} \! \int\! \frac{d^4k}{(2\pi)^4}\ Z_A \Delta_{\mu\nu}(q)  \gamma_\mu\, Z_2\, S(k)\,  \frac{\Gamma_{\nu} (k,p)}{Z_1} \ .
\end{equation}
\end{spreadlines}
After simplification one arrives at the renormalized DSE~\eqref{DSEquark}, where the integral is multiplied by the $\mu$-dependent vertex renormalization constant
$Z_1$.  Now, recall that the STI  for the quark-gluon vertex~\eqref{STI}  imposes the all-order constraint on the renormalization  factors,
\begin{equation}
   Z_1 = \frac{Z_H Z_2}{Z_c} \ ,
 \label{STIconstantes}
\end{equation}
in which $Z_H$, defined by $H (k,p,\mu^2) = Z_H(\mu^2) H_0(k,p)$, is the renormalization of the quark-ghost kernel discussed in Section~\ref{quarkghost} and $Z_c$ is the
renormalization constant of the ghost propagator~\eqref{ghostprop}:
\begin{equation}
    G(q^2,\mu^2 ) = Z_c^{-1} (\mu^2 ) G_0(q^2) \ .
\end{equation}
It is known that the ghost-gluon vertex is not ultraviolet divergent in Landau gauge and that the quark-ghost kernel is finite at one-loop order~\cite{Taylor:1971ff}.
This also is the case in the  dressed-perturbative approach~\cite{Aguilar:2010cn} we employ and we choose $Z_H =1$. The STI condition~\eqref{STIconstantes} thus
simplifies to~\cite{Marciano:1977su},
\begin{equation}
   Z_1 = \frac{Z_2}{Z_c} \ ,
\label{z1z2zc}
\end{equation}
which is still not the relation $Z_1 = Z_2$ imposed by the WFGTI in QED. Since the self-energy  $\Sigma (p^2)$ is also finite at one loop in Landau gauge, the authors
of Ref.~\cite{Aguilar:2010cn} set $Z_2=1$ which simplifies the STI condition further to $Z_1 = 1/Z_c$, and in order to insure multiplicative renormalizability the replacement
$Z_c^{-1} \to G(q^2)$ was advocated. Similarly, also insisting on the renormalization-point independence of $M(p^2)$, the Curtis-Pennington vertex~\cite{Curtis:1990zs} is
multiplied by a factor of $G(q^2)$ in a non-Abelian ansatz for the quark-gluon vertex~\cite{Fischer:2003rp}.

An intuitive motivation for this substitution is provided by the observation that the product $g^2 \Delta(q^2) G^2(q^2)$ is a renormalization group invariant quantity in the
Taylor  scheme\footnote{\,This corresponds to the Taylor kinematics of vanishing incoming
ghost momentum in the MOM scheme~\cite{Boucaud:2008gn}. The MOM scheme itself is defined by setting the scalar coefficient function of the Green function
to its tree-level value in a specific kinematic configuration.}
and can be used to define an effective strong coupling in the DSE kernel. As argued in Ref.~\cite{Aguilar:2010cn}, the prescription $Z_c^{-1} \to G(q^2)$ compensates for the missing transverse part of the  quark-gluon vertex in
their approach. This is because omitting the transverse vertex components leads to mishandling of overlapping divergences and therefore compromises the
multiplicative renormalizability of the DSE. Indeed, we demonstrated in Ref.~\cite{Albino:2018ncl} that the transverse vertex is crucial to satisfy multiplicative
renormalizability and in here  \emph{we include all components of the vertex\/}.

If we solved simultaneously the coupled quark, gluon and ghost DSE, we would know the value of $Z_c$. This is not our task here, as we pursue the hybrid approach of
solving the quark DSE~\eqref{DSEquark} using as input  renormalized gluon and ghost propagators from lattice QCD. Note that the STI and TSTI to be discussed in
Sections~\ref{generaltensor} and \ref{TSTIs} involve the ghost propagator and the quark-ghost scattering amplitude. Consequently, the vertex  $\Gamma_\mu^a (k, p)$
we derive therefrom also depends on their scalar dressing functions. We therefore absorb $Z_c$ on the right-hand side of Eq.~\eqref{z1z2zc} in the ghost-dressing
function $G (q^2)$ (see Eqs.~\eqref{lambda1QCD} to \eqref{tau8QCD}) or, similarly, we may absorb $Z_c$ in the strong coupling $\alpha_s =g^2/4\pi$
which is an overall factor in front of the integral. As we allow for a certain variation of  $\alpha_s(\mu^2)$ at a given renormalization scale imposed by lattice QCD,
the impact of this simplification is minimal. Thus, the quark DSE we are concerned with  in this work is,
\begin{equation}
      S^{-1}(p)  = \, Z_2 \,  i\, \gamma\cdot p + Z_4 \, m
                      + \, Z_2\, \frac{16\pi \alpha_s}{3} \int^\Lambda\!\!  \frac{d^4k}{(2\pi)^4}\  \Delta_{\mu\nu} (q)\, \gamma_\mu \,  S(k)\,\Gamma_\nu (k,p) \ ,
\label{finalDSEquark}
\end{equation}
with all the 12 tensor structures of the fully dressed quark-gluon vertex we derive in the next sections.


\subsection{The quark-gluon vertex: general tensor structure \label{generaltensor} }

As emphasized in Section~\ref{DSE}, the dressed quark-gluon vertex is a fundamental ingredient in the QCD gap equation and its contributions in the infrared domain play
a pivotal role for the emergence of a constituent-quark mass. The vertex will result in strong DCSB if at least one form factor associated with its tensor structure provides
sufficient enhancement in the infrared domain; this feature is commonly associated with the rainbow-ladder truncation of the gap- and bound-state equations. However, as
argued for example in Refs.~\cite{Bashir:2012fs,ElBennich:2012tp,Rojas:2014aka,Mojica:2017tvh}, this simplest ansatz has limitations when applied to radially excited mesons,
the axialvector channel and heavy-flavored mesons. It is therefore of crucial importance to understand the individual as well as collective contributions of the full vertex to
the gap equation.

\begin{figure}[t!]
\vspace*{-1cm}
\centering
  \includegraphics[scale=0.45]{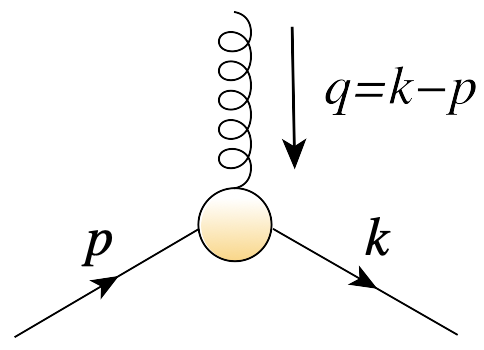}
  \caption{Diagrammatic representation of the dressed quark-gluon vertex $\Gamma_\mu (k,p)$ with its momentum flow defined by the incoming quark momentum $p$,
                the outgoing momentum $k$ and $q=k-p$.    }
\label{quark-gluon-vertex}
\end{figure}

To begin with, as in QED, the fermion-gauge-boson vertex in QCD satisfies a set of constraints imposed by gauge symmetry, multiplicative renormalizability,  LKFT as well as the
$C$, $P$ and $T$ symmetry properties of the bare vertex; see, for instance, the discussions in Refs.~\cite{Bashir:2004mu,Bermudez:2017bpx}. Therefore, in a general though
not unique decomposition, the fermion-gauge-boson vertex can be written in terms of twelve linearly independent structures~\cite{Ball:1980ay}.  In QCD, the STI
associated with this vertex  is given by,
\begin{equation}
   i\, q \cdot \Gamma^a ( k, p )  =   G (q^2) \Big [ S^{-1} ( k) \, H^a( k, p )  -    \bar  H^a (p, k) S^{-1}(p) \Big ] \ ,
\label{STI}
\end{equation}
where $q=k-p$ and the momentum flow is depicted in Figure~\ref{quark-gluon-vertex} and $S^{-1} (p) = i A(p^2) \, \gamma \cdot p +B(p^2)$ is the inverse of the propagator 
in Eq.~\ref{DEsol}.  The color index $a$ denotes again the implicit SU(3) group generators. The ghost-dressing function $G(q^2)$ is defined by the ghost propagator,
\begin{equation}
  D^{a b} (q^2) = -\, \delta^{ab} \, \frac{G(q^2)}{q^2} \ ,
 \label{ghostprop}
\end{equation}
renormalized as $G(\mu^2) = 1 $,  $H^a( k, p ) = H ( k, p ) t^a$ is the quark-ghost scattering amplitude and $\bar H^a( p,k )$ is obtained from the conjugate, $H^\dagger(-p,-k)$,
and the  momentum exchange $k \leftrightarrow - p$. This identity allows us to determine the four components of the vertex which are not transverse to the gluon momentum $q$. 
However, this doesn't exhaust the number of independent  tensor structures that can be formed from the momenta $k$ and $p$  and the Dirac matrices. The full vertex,
$\Gamma_\mu^a (k,p)  = \Gamma_{\mu}(k,p)\, t^a$ contains  additional transverse vertex components  and can be decomposed as the sum~\cite{Ball:1980ay},
\begin{equation}
    \Gamma_{\mu}(k,p)  = \Gamma^{L}_{\mu}(k,p)  + \Gamma^{T}_{\mu}(k,p) \ .
\label{BallChiu}
\end{equation}
The transverse vertex $\Gamma^T_{\mu}(k,p) $ vertex in Eq.~\eqref{BallChiu} is thus naturally defined by,
\begin{equation}
 i\,  q \cdot \Gamma^T (k,p)  =  0 \, .
 \label{t-vertexdef}
\end{equation}
The STI~\eqref{STI} constrains the so-called longitudinal vertex, $\Gamma_{L}^{\mu}(k,p)$, to four independent structures,
\begin{eqnarray}
 \Gamma^{L}_{\mu} (k,p)   =  \lambda_1 (k,p)  \,  \gamma_\mu + \lambda_2 (k,p)  \,   \tfrac{1}{2}\,  t_\mu\,  \gamma \cdot t
                                          -  i  \lambda_3 (k,p)   \, t_\mu -   \lambda_4 (k,p)  \,  \sigma_{\mu\nu} t_\nu \,,   \hspace*{4mm}
\label{LVD}
\end{eqnarray}
where $t=p +k$ is the sum of the incoming and outgoing quark momenta.

The transverse vertex, ultraviolet-finite at one-loop order~\cite{Davydychev:2000rt,Fernandez-Rangel:2016zac}, can generally be decomposed into eight independent
 tensor structures multiplied by associated scalar form factors, $\tau_{i}(k ,p)$~\cite{Ball:1980ay}:
\begin{equation}
    \Gamma^T_\mu (k,p) =  \sum_{i=1}^{8} \tau_{i}(k,p) \, T^i_\mu (k,p)\, .
\label{t-vertexstruc}
\end{equation}
For the kinematical configuration in Figure~\ref{quark-gluon-vertex}, the tensors are defined as:
\begin{eqnarray}
T^{1}_{\mu}(k,p) &=& i \left[ p_{\mu} (k \cdot q) -k_{\mu} (p \cdot q) \right] \,,  
\label{tbdef1}  \\   [5pt]
T^{2}_{\mu}(k,p) &=& \left [ p_{\mu} (k \cdot q) -k_{\mu} (p \cdot q) \right] \gamma \cdot t \,, \\ [5pt]
T^{3}_{\mu}(k,p) &=& q^{2} \gamma_{\mu} - q_{\mu}\, \gamma \cdot q \,,  \\ [5pt]
T^{4}_{\mu}(k,p) &=&  i q^{2}\left[\gamma_{\mu} \gamma \cdot t-t_{\mu}\right]+2 q_{\mu} p_{\nu} k_{\rho} \sigma_{\nu \rho} \, , \\ [5pt]
%
T^{5}_{\mu}(k,p) &=& \sigma_{\mu \, \nu} q_{\nu}  \,,  \\  [5pt]
T^{6}_{\mu}(k,p) &=& - \gamma_{\mu} \left( k^{2}-p^{2} \right) + t_{\mu}\, \gamma \cdot q \,,  \\  [5pt]
T^{7}_{\mu}(k,p) &=&  \tfrac{i}{2} (k^{2}-p^{2}) \left[ \gamma_{\mu} \gamma \cdot t - t_{\mu} \right] + t_{\mu}\, p_{\nu} k_{\rho} \sigma_{\nu  \rho}   \,,  \\ [5pt]
T^{8}_{\mu}(k,p) &=& -i \gamma_{\mu} \, p_{\nu} k_{\rho} \sigma_{\nu  \rho} - p_{\mu} \, \gamma \cdot k + k_{\mu}\, \gamma \cdot p \, .
\label{tbdef8}
\hspace*{1.2cm}
\end{eqnarray}
Here, we adopt a slightly modified tensor basis~\cite{Kizilersu:1995iz} with respect to the one derived by Ball and Chiu~\cite{Ball:1980ay}. This ensures all
transverse form factors are independent of any kinematic singularities in one-loop perturbation theory in an arbitrary covariant gauge.

As discussed in more detail in Refs.~\cite{Bashir:2004mu,Bermudez:2017bpx}, the fully dressed vertex must exhibit the same properties under charge-conjugation
transformation as the bare vertex, which constrains all the $\tau_i(k,p) $ in Eq.~\eqref{t-vertexstruc} to be symmetric under the momentum interchange
$k \leftrightarrow p$, with the exception of the odd functions $\tau_{4}(k,p)$ and $\tau_{6}(k,p)$:
\begin{eqnarray}
\label{tausymmetry1}
   \tau_{i}(k,p)&=& \tau_{i}(p,k) \,, \quad i=1,2,3,5,7,8  \\
    \tau_i (k,p)&=& -\tau_i (p,k) \, , \quad i=4,6 \ .
\label{tausymmetry2}
\end{eqnarray}
Similarly, $\lambda_1(k,p) $, $\lambda_2 (k,p) $, $\lambda_3 (k,p) $ and $\lambda_4 (k,p) $ are symmetric under $k \leftrightarrow p$,
so as to preserve the charge-conjugation properties of the bare vertex.

The STI~\eqref{STI} completely determines the form factors in Eqs.~\eqref{LVD} and their expressions are given in In Section~\ref{TSTIs}. The transverse scalar functions
in Eq.~\eqref{t-vertexstruc}, on the other hand, remain undetermined. Nonetheless, as in QED, two additional transverse identities allow to analogously constrain the form
of these transverse vertex functions.


\subsection{Transverse Slavnov-Taylor identities  \label{TSTIs}  }

A complete fermion-boson vertex  based on gauge techniques, such as the one derived from the TTI~\cite{Qin:2013mta,Albino:2018ncl}, is a first step towards
a realistic quark-gluon interaction in QCD valid in any coupling regime. As in QED, gauge invariance as well as Lorentz and charge transformation invariance
constrain the tensor structure of the fully dressed three-point function.

As we already saw, the corresponding gauge symmetry preserving relation between Green functions  in QCD theory is provided by the STI of Eq.~\eqref{STI},
which relates the quark-gluon vertex with the ghost-dressing function, the inverse quark propagator and the quark-ghost scattering kernel, $H( k,p)$, and its
conjugate, $\bar  H (k,p )$.  The non-transverse vertex saturates this STI while the transverse vertex is automatically  conserved. As in QED, this allows
to decompose the fermion-boson vertex into a so-called longitudinal and a transverse part with a tensor basis for which we  choose the one defined in
Eq.~\eqref{LVD} and Eqs~\eqref{tbdef1}--\eqref{tbdef8}.  On the other hand, the STI  is a consequence of SU(3) gauge invariance and merely
makes a statement about the non-transverse components of the vertex. By itself, it fails  to ensure gauge covariance in its totality. Under certain simplifying
conditions, the WFGTI can be recovered form the STI which leads to an abelianized form of the longitudinal quark-gluon vertex~\cite{Aguilar:2010cn,Rojas:2014tya}.

How do the DSE solutions of propagators and vertices depend on an arbitrary gauge transformation?  In QED the answer to this question can be obtained
from the well-defined set of LKFT laws which leave the DSE and WGTI form-invariant. Given a suitable ansatz for the three-point Green function, $\Gamma(k,p)$,
that is invariant under a LKFT to another gauge, one can ensure gauge covariance and the WFGTI is satisfied in either gauge. The general rules that govern
the LKFT, though, are far from trivial and an extension to the quark propagator in any gauge has only recently been established~\cite{Aslam:2015nia}.

In analogy to QED, the transverse components of the vertex can be constrained by TSTI that relate to the curl of the quark-gluon vertex. Their Dirac
structure is identical to that of the TTI and in addition they involve the ghost-dressing function and the quark-ghost scattering kernel. The TSTI can be
derived in a functional approach~\cite{He:2009sj} and read:
\begin{eqnarray}
 q_{\mu} \Gamma_{\nu}^a(k,p) -q_{\nu} \Gamma_{\mu}^a (k,p)
      & =&  G(q^2) \left [ S^{-1}(p) \sigma_{\mu\nu}\, H^a( k, p)  + \bar  H^a (p, k)\,  \sigma_{\mu\nu} S^{-1}(k) \right ] \nonumber \\
      & +  &  2 i m\, \Gamma_{\mu\nu}^a (k,p) + t_{\alpha} \epsilon_{\alpha \mu \nu \beta}\, \Gamma^{5a}_{\beta}(k,p) +  A^{a}_{\mu\nu}(k,p) \,,
 \label{TSTI-Vector}   \\  [12pt]
  q_{\mu} \Gamma^{5a}_{\nu}(k,p) - q_{\nu} \Gamma^{5a}_{\mu}(k,p)
      & = &  G(q^2) \left [  S^{-1}(p) \sigma^{5}_{\mu\nu} \, H^a( k, p)  - \bar  H^a (p, k) \, \sigma^{5}_{\mu\nu} S^{-1}(k) \right ] \nonumber \\
      & + & t_{\alpha} \epsilon_{\alpha \mu \nu \beta} \, \Gamma_{\beta}^a (k,p) + V^{a}_{\mu\nu}(k,p) \, ,
\label{TSTI-Axial}
\end{eqnarray}
where $m$ is the renormalized current-quark mass in the DSE, $\sigma^{5}_{\mu\nu}=\gamma_{5} \sigma_{\mu\nu}$, and $\Gamma_{\mu\nu}^a (k,p)$ is an inhomogeneous
tensor vertex.  The two tensor structures, $A^a_{\mu\nu}$ and $V^a_{\mu\nu}$, are Fourier transforms of four-point functions in coordinate space which involve a Wilson line
to preserve gauge invariance~\cite{He:2009sj}. Moreover, $\Gamma_{\mu}^a(k,p)$ couples to the axialvector vertex via the fourth term on the right-hand side of 
Eq.~\eqref{TSTI-Vector} and likewise,  $\Gamma^{5a}_{\mu}(k,p)$ couples to the vector vertex.  We stress that these identities bear no explicit dependence on the covariant
 gauge parameter. They are valid for any covariant gauge and are not limited to the perturbative case.

We follow Refs.~\cite{Qin:2013mta,Albino:2018ncl} and proceed in analogous manner to decouple Eqs.~\eqref{TSTI-Vector} and \eqref{TSTI-Axial}.
 This is done so by introducing the two tensors,
\begin{eqnarray}
   T^{1}_{\mu\nu} & = & \frac{1}{2} \, \epsilon_{\alpha \mu \nu \beta} t^{\alpha} q^{\beta} \,,
\label{T1contract}   \\  [10pt]
   T^{2}_{\mu\nu} & = & \frac{1}{2} \, \epsilon_{\alpha \mu \nu \beta}
\gamma^{\alpha} q^{\beta} \, .
\label{T2contract}
\end{eqnarray}
Contracting the axialvector identity~\eqref{TSTI-Axial} with the tensors \eqref{T1contract} and \eqref{T2contract} yields zero on the left-hand side of the equation,
whereas operating the contractions on the right-hand sides with the identities,
\begin{eqnarray}
T^{1}_{\mu\nu}\, t_{\alpha} \epsilon_{\alpha \mu \nu \beta} \ \Gamma_{\beta}(k,p) &=& t^{2} q \cdot \Gamma(k,p) - q \cdot t \, t \cdot \Gamma(k,p) \,  ,  \\  [10pt]
T^{2}_{\mu\nu}\,  t_{\alpha} \epsilon_{\alpha \mu \nu \beta} \ \Gamma_{\beta}(k,p) &=& \gamma \cdot t \,  q \cdot \Gamma(k,p) - q \cdot t  \, \gamma \cdot \Gamma(k,p) \, ,
\end{eqnarray}
one can recast the two contractions of the axialvector identity into the new form,
\begin{eqnarray}
    q \cdot t\,  t \cdot \Gamma(k,p) &  =  & G(q^2)\, T^{1}_{\mu\nu} \left[  S^{-1}(p) \sigma^{5}_{\mu \nu}  \, H( k, p)  -  \bar H( p, k)  \sigma^{5}_{\mu \nu} S^{-1}(k)  \right  ]
\nonumber \\[6pt]
                     &  +  & t^{2} q \cdot \Gamma(k,p) + T^{1}_{\mu\nu} V_{\mu\nu}  \ ,
    \label{TWFGTI1}    \\  [10pt]
    q \cdot t \, \gamma \cdot \Gamma(k,p) & = & G(q^2)\, T^{2}_{\mu\nu} \left[ S^{-1}(p) \sigma^{5}_{\mu \nu} \, H( k, p)  -  \bar H( p, k)  \sigma^{5}_{\mu \nu} S^{-1}(k)  \right ]
\nonumber \\   [6pt]
                                                                 & + &  \gamma \cdot t\, q \cdot \Gamma(k,p) + T^{2}_{\mu\nu} V_{\mu\nu} \ ,
   \label{TWFGTI2}
\end{eqnarray}
in which we dropped color indices for simplicity's sake. Remarkably, these two new identities involve only the vector vertex, $\Gamma_{\mu}(k,p)$, and do not explicitly
depend on the quark mass $m$. Likewise, information about the axialvector vertex, $\Gamma^5_{\mu}(k,p)$, can be obtained through an analogous procedure involving
the vector TSTI~\eqref{TSTI-Vector}.

We have thus uncoupled the vector from the axialvector vertex and obtain, in analogy with Ref.~\cite{Qin:2013mta}, three matrix-valued equations
for the scalar projections of $\Gamma_\mu (k,p)$, namely  Eqs.~\eqref{STI}, \eqref{TWFGTI1}  and \eqref{TWFGTI2}.
They form a set of twelve linearly independent and coupled linear equations for the twelve unknown scalar vertex functions, $\lambda_i (k,p)$ and $\tau_i (k,p)$,
which can be solved exactly provided $H^a(k,p)$,  $\bar H^a(p,k)$  and $V^{a}_{\mu\nu}(k,p)$ are known. The terms $T^{1}_{\mu\nu} V_{\mu\nu}$
and $T^{2}_{\mu\nu} V_{\mu\nu}$ are unknown objects, yet they are Lorentz-scalar objects which, without loss of generality, can be conveniently expressed as,
\begin{spreadlines}{8pt}
\begin{align}
   i\, T^{1}_{\mu\nu} V_{\mu\nu}  &=  Y_{1}(k,p)  \mathbbm{1}_{D} +   Y_{2}(k,p)   \gamma \cdot q
                       +  Y_{3}(k,p)  \gamma \cdot t  +  Y_{4}(k,p)\left[ \gamma \cdot q, \gamma \cdot t \, \right  ]  \, , \label{TV1} \\
   i\, T^{2}_{\mu\nu} V_{\mu\nu}  &=   Y_{5}(k,p)  \mathbbm{1}_{D} +   Y_{6}(k,p)   \gamma \cdot q
                       +  Y_{7}(k,p)  \gamma \cdot t  +  Y_{8}(k,p)\left[ \gamma \cdot q, \gamma \cdot t \, \right  ]  \, , \label{TV2}
\end{align}
\end{spreadlines}
where $Y_{i}(k,p)$ are hitherto unconstrained scalar functions. 


A final ingredient we require to derive the expressions for the form factors $\lambda_{i}(k,p)$ and $\tau_{i}(k,p)$ is the quark-ghost scattering amplitude,
$H^a(k,p) = H(k,p) t^a$, which so far has been studied in a dressed-perturbative approach and will be discussed in more detail in Section~\ref{quarkghost}.
Nonetheless, employing a similar expedient as in Eqs.~\eqref{TV1} and \eqref{TV2}, we can decompose the matrix-valued kernel and parameterize
it as follows:
\begin{spreadlines}{8pt}
\begin{align}
        H( k, p ) &=   X_0( k, p ) \mathbbm{1}_{D} +  X_1( k, p )  \,\gamma\cdot k
          +  X_2 ( k, p ) \,   \gamma\cdot  p +  X_3( k, p )\, \left[ \gamma \cdot k, \gamma \cdot p \, \right]  , \label{Xidef1}        \\
\
 \bar  H( p, k ) &=  X_0( p, k)\mathbbm{1}_{D} -  X_2( p, k )\,  \gamma\cdot  k
         -  X_1( p, k) \,  \gamma\cdot  p +  X_3( p, k ) \, \left[ \gamma \cdot k, \gamma \cdot p \,\right] \,  .  \label{Xidef2}
\end{align}
\end{spreadlines}
Perturbative expressions for the form factors $X_i (k,p)$ are known to one-loop order~\cite{Davydychev:2000rt} and yield $X_0  (k,p) = 1 + \mathcal{O}(g^2)$
and $X_i = \mathcal{O}(g^2)$, $i = 1, 2, 3$.  The same diagrammatic approach based on dressed propagators and vertices that incorporate all $X_i (k,p)$ form 
factors in the longitudinal vertex appears to mostly corroborate this dominance, while $X_3(k,p)$ may become sizable~\cite{Aguilar:2016lbe} in certain
kinematic limits. For the remainder of this work, we make the simplifying assumption that the contribution of $X_0(k,p)$ dominates and limit ourselves
to the kinematic configuration, $k=q/2$, $p=- q/2 \, \Rightarrow\,  H( k, p )  = H(q/2,- q/2)$, and therefore $X_0 ( k,p) \approx  X_0 ( q^2)$. We neglect
sub-leading form factors as their contributions are negligibly small compared to the strength of the transverse vertex, as will be shown shortly.

The decomposition in terms of Lorentz covariants of Eqs.~\eqref{TV1}, \eqref{TV2}, \eqref{Xidef1} and \eqref{Xidef2} allows us to write the  identities~\eqref{STI}, \eqref{TWFGTI1} 
and \eqref{TWFGTI2} as a matrix-valued equation system that can be solved by different means. With a set of four projections, with respect to the Dirac trace, one obtains
four linearly independent, coupled linear equations whose solutions yield the $\lambda_i (k,p)$ form factors  and, likewise, the $\tau_i(k,p)$ can be isolated using 
eight different projections.  Thus, with the four projections of Eq.~\eqref{STI} we arrive the following scalar vertex  functions of the non-transverse vertex:
\begin{eqnarray}
  \lambda_1 (k,p)  & = & \tfrac{1}{2}\,  G(q^2) X_0 (q^2)\left[ A (k^2) + A(p^2)   \right] \ ,
 \label{lambda1QCD}   \\ [5pt]
   \lambda_2 (k,p)  & = &  G(q^2) X_0 (q^2)\, \frac{ A (k^2) - A(p^2)}{k^2-p^2}  \ ,
 \label{lambda2QCD}   \\ [5pt]
  \lambda_3 (k,p)  & = &   G(q^2) X_0 (q^2)\, \frac{ B (k^2) - B(p^2)}{k^2-p^2}  \ ,
 \label{lambda3QCD}   \\ [5pt]
  \lambda_4 (k,p)  & = &\ 0 \ .
 \label{lambda4QCD}
\end{eqnarray}
Note that in the absence of the sub-leading $X_{1,2,3}(k,p)$ form factors, $\lambda_4 (k,p)  =0$, as in QED~\cite{Ball:1980ay}.
The transverse vertex functions are derived from four projections of each, Eqs.~\eqref{TWFGTI1} and \eqref{TWFGTI2}:
\begin{eqnarray}
   \tau_1(k,p) & = & - \frac{ Y_{1} }{ 2 (k^{2} - p^{2}) \nabla(k,p) } \ ,
\label{tau1QCD}   \\ [7pt]
   \tau_2(k,p) &=& - \frac{Y_{5} - 3 Y_{3}}{ 4 (k^{2} - p^{2}) \nabla(k,p) } \ ,
\label{tau2QCD}  \\ [7pt]
   \tau_3(k,p) &=&  \frac{1}{2}\,  G(q^2) X_0 (q^2)\left[ \frac{ A (k^2) - A(p^2)}{k^2-p^2}  \right] +   \frac{Y_{2}}{4\nabla(k,p)}
   - \frac{ t^{2} (Y_{3} - Y_{5}) }{ 8(k^{2} - p^{2}) \nabla(k,p) } \ ,    \hspace*{1cm}
\label{tau3QCD}  
\end{eqnarray}
\begin{eqnarray}
   \tau_4(k,p) &=&  - \frac{ 6 Y_{4} + Y_{6}^{A} }{8\nabla(k,p) } - \frac{t^{2} Y_{7}^{S} }{ 8(k^{2} - p^{2}) \nabla(k,p) } \  ,
\label{tau4QCD} \\  [7pt]
   \tau_5(k,p) &=& -  G(q^2) X_0 (q^2)\left[ \frac{ B (k^2) - B(p^2)}{k^2-p^2}  \right ]   - \frac{2 Y_4 + Y_6^A}{2 (k^{2}-p^{2})} \ ,
\label{tau5QCD}  \\ [7pt]
   \tau_6(k,p) &=& \frac{(k-p)^{2} Y_{2} }{ 4 (k^{2} - p^{2}) \nabla(k,p) } -\frac{Y_3 - Y_5}{8\nabla(k,p) }  \ ,
\label{tau6QCD}  \\ [7pt]
   \tau_7(k,p) &=& \frac{q^2 (6Y_4 +Y_6^A)}{4(k^2-p^2)\nabla(k,p)}   +\frac{Y_7^S}{4\nabla(k,p )} \ ,
\label{tau7QCD}  \\ [7pt]
  \tau_8(k,p) &=&  - G(q^2) X_0 (q^2)\left[ \frac{ A (k^2) - A(p^2)}{k^2-p^2}  \right] - \frac{2 Y_8^A}{k^{2}-p^{2}}  \ ,
\label{tau8QCD}
\end{eqnarray}
where $Y_{i} \equiv Y_{i}(k,p)$ and the Gram determinant is defined by:
\begin{equation}
     \nabla (k,p) = k^{2} p^{2} - (k \cdot p)^{2} \,  .
     \label{Gram}
\end{equation}
In addition, the vertex transformation properties under charge conjugation determine the symmetry properties
of the $Y$-functions:
\begin{eqnarray}
      Y_{i}(k,p) & =  & Y_{i}(p,k) \ , \qquad  i=2,6^{S},7^{S},8^{S}\ ,
\label{Y-functions symmetric properties}    \\
      Y_{i}(k,p)  & = &  -Y_{i}(p,k) \ , \quad i=1,3,4,5,6^{A},7^{A},8^{A} \ ,
      \label{Y-functions antisymmetric properties}
\end{eqnarray}
where, as in Ref.~\cite{Albino:2018ncl}, we introduce the decomposition,
\begin{equation}
    Y_{i}(k,p) = Y_{i}^{S}(k,p) + Y^{A}_{i}(k,p) \ ,
  \label{Y8symmetric-antisymmetric definition}
\end{equation}
for $i=6,7,8$, where the superscripts $S$ and $A$ denote the symmetric and antisymmetric parts of the corresponding $Y_{i}$ functions for $k \leftrightarrow p$.
Note that there is no contribution of $Y_{6}^{S}$, $Y_{7}^{A}$ and $Y_{8}^{S}$ to the form factors $\tau_i(k,p)$ in Eqs.~\eqref{tau1QCD}
to \eqref{tau8QCD} as a consequence of the  symmetry properties in  Eqs.~\eqref{tausymmetry1} and \eqref{tausymmetry2} which imply:
\begin{eqnarray}
Y_{6}^{S}(k,p)   &=& - \frac{(k^{2}-p^{2})  Y_{1}(k,p)}{4 \nabla (k,p)} \ , \label{Y6 sym}   \\ [10pt]
Y_{7}^{A}(k,p)  &=& \frac{ q^{2}\, Y_{1}(k,p)}{4 \nabla (k,p)} \ , \label{Y7antisym}  \\ [10pt]
Y_{8}^{S}(k,p)  & = &  -\frac{ (k-p)^{2} \, Y_{2}(k,p)}{ 8 \nabla (k,p) }   +   \frac{(k^{2}- p^{2}) Y_{3}(k,p) }{ 8 \nabla (k,p) } \  .
\label{Y8sym}
\end{eqnarray}

We stress that while Eqs.~\eqref{lambda1QCD} to  \eqref{tau8QCD} are analytic expressions, the remaining challenge stems from our ignorance of the scalar
$Y_i(k,p)$ functions. We shall constrain their analytical form with an ansatz for the transverse vertex in perturbative QCD~\cite{Bashir:2011dp} in
Section~\ref{Yi-ansatz} and postpone the actual calculation of the nontrivial four-point function $V^{a}_{\mu\nu}(k,p)$.


\subsection{Quark-ghost scattering amplitude \label{quarkghost} }

\begin{figure}[t!]
\vspace*{-3mm}
\centering
  \includegraphics[scale=0.4]{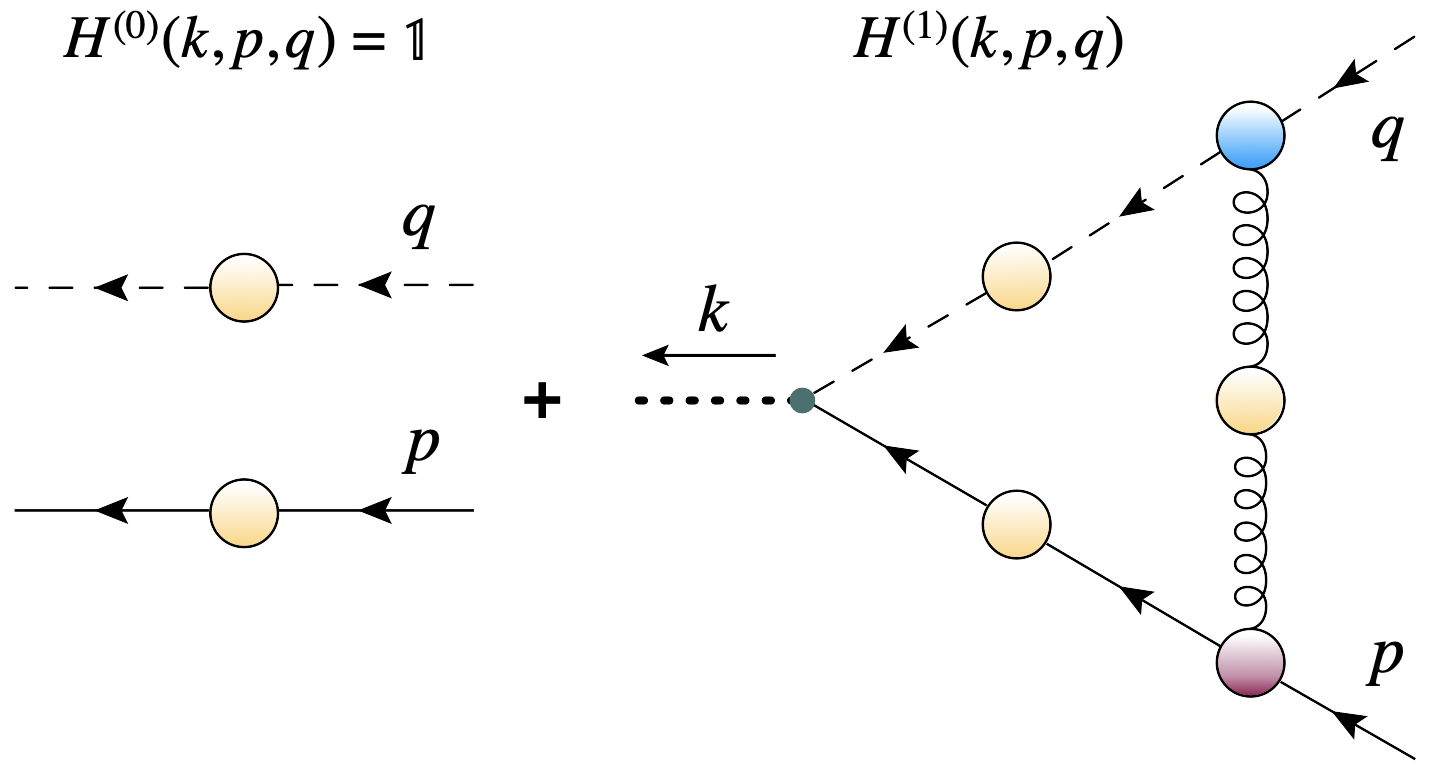}
  \vspace*{2mm}
  \caption{Diagrammatic representation of the unconnected and connected quark-ghost scattering amplitude in its dressed one-loop approximation with the simplified kinematic
              choice $k=q/2$ and $ p=-q/2$. The yellow-shaded circles denote the dressed quark, gluon and ghost propagators, while the blue- and purple-shaded circles represent the
                dressed ghost-gluon and quark-gluon vertices, respectively. The outgoing quark and ghost propagators are joined in a non-standard ``vertex''.}
\label{H-diagram}
\end{figure}

The nonperturbative behavior of the quark-ghost scattering kernel $H^a(k,p)$ introduced in the STI~\eqref{STI} is unknown. A tractable ansatz that allows for a connection
with the perturbative result~\cite{Davydychev:2000rt} is given by the one-loop dressed approximation to $H^a(k,p)$~\cite{Aguilar:2010cn}, whose diagrammatic representation
is depicted in Figure~\ref{H-diagram}. We use the tensorial decomposition of Eqs.~\eqref{Xidef1} and \eqref{Xidef2} and, as mentioned previously, limit ourself to the
dominant form factor $X_0(k,p)$ with the momentum configuration $k=q/2$ and  $p = -q/2$, so that $X_0(q/2,-q/2) \equiv X_0(q^2)$ does not depend on angles. Applying 
the kinematics of the Feynman diagram in Figure~\ref{H-diagram}, we project out $X_0(q^2)$,
\begin{align}
 X_0(q^2)  & = \,   \tfrac{1}{4} \operatorname{Tr}_{\mathrm{CD}} H(q/ 2,-q / 2,q)            \nonumber \\
                 & =  \,  1 + \frac{C_A}{4} \, g^2 \! \int^\Lambda\!\!   \frac{d^4 \ell}{(2\pi)^4}\ \Delta_{\mu \nu}(\ell)  D( \ell+ q )
                          \operatorname{Tr}_D \big [  G_{\nu}\,  S\left (\ell + \tfrac{q}{2} \right ) \Gamma_{\mu} \left (\ell+\tfrac{q}{2} , -\tfrac{q}{2} \right ) \! \big ] \ ,
  \label{X0int}
\end{align}
where $C_A= 3$ is the Casimir operator in the adjoint representation,
\begin{equation}
    G_{\nu} = i \left (\ell_\nu +  q_\nu \right )  H_1  -  i \ell_\nu\,  H_2  \ ,
\label{ghostgluonvertex}
\end{equation}
is the dressed ghost-gluon vertex and $\ell+q$ and $q$ are respectively the outgoing and incoming ghost momenta, while $\ell$ is the gluon momentum exchanged
between the quark and the ghost. Note that that the terms proportional to $\ell_\mu$ in the vertex~\eqref{ghostgluonvertex} are transverse to the gluon
propagator $\Delta_{\mu \nu}(\ell)$ in Landau gauge; $H_2$ does therefore not  contribute to the quark-ghost kernel. Moreover, the momentum $q$ coincides with
the gluon  momentum in the gap equation~\eqref{DSEquark}.

The remaining form factor $H_1 (\ell+q,q) \simeq H_1 (\ell+q ,0 ) \equiv H_1(x)$  is parametrized by the expression~\cite{Dudal:2012zx},
\begin{equation}
       H_1 (x) =  c \left( 1 + \frac{a^2 x^2}{x^4 + b^4} \right) + (1 - c)\, \frac{w^4}{w^4 + x^4} \ ,
 \label{H1}
\end{equation}
fitted to lattice-QCD data: $c = 1.26$, $a = 0.80$~GeV, $b = 1.3$~GeV and $w = 0.65$~GeV. In Ref.~\cite{Rojas:2013tza} the replacement of the tree-level ghost-gluon
vertex by a dressed vertex enhanced $X_0(q^2)$ overall and by about 20\% in the peak region $q^2=$1--4~GeV$^2$. Adding the transverse part of the vertex in the gap 
equation this observation is not significantly altered. Note that $\Gamma_\mu(k,p) =  \Gamma^{L}_{\mu}(k,p)  + \Gamma^{T}_{\mu}(k,p)$
enters both the DSE~\eqref{finalDSEquark} and the quark-ghost scattering amplitude in Eq.~\eqref{X0int} for consistency.


\subsection{Ansatz to constrain the nonlocal Lorentz scalars $T^{1}_{\mu \nu} V_{\mu \nu}$ and $T^{2}_{\mu \nu} V_{\mu \nu}$  \label{Yi-ansatz} }

The vertex ansatz proposed in Ref.~\cite{Bashir:2011dp}  is constrained by two requirements: it provides the multiplicative renormalizability of the fermion
propagator and produces a gauge-independent critical coupling for DCSB. The appeal of this ansatz lies in the fact that it is merely expressed in terms of the
quark propagator's scalar and vector functions, $B(p^{2})$ and $A (p^{2})$, respectively. In the particular kinematical configuration $k^2 \gg p^2$, the transverse
form factors of this ansatz are given by:
\begin{eqnarray}
   \tau_{1}(k,p) &=& a_{1} \frac{ B(k^{2}) - B(p^{2}) }{ (k^{2}+p^{2}) (k^{2}-p^{2}) } \,, \label{Rociotransverse1}   \\ [6pt]
  \tau_{2}(k,p) &=& a_{2} \frac{ A(k^{2}) - A(p^{2}) }{ (k^{2}+p^{2}) (k^{2}-p^{2}) } \,, \\ [6pt]
  \tau_{3}(k,p) &=& a_{3} \frac{ A(k^{2}) - A(p^{2}) }{ k^{2}-p^{2} } \, ,   \\ [6pt]
  \tau_{4}(k,p) &=& a_{4} \frac{ B(k^{2}) - B(p^{2}) }{4 k^{2}p^{2} } \,, \\ [6pt]
  \tau_{5}(k,p) &=& -a_{5} \frac{ B(k^{2}) - B(p^{2}) }{ k^{2}-p^{2} } \,,  \\ [6pt]
  \tau_{6}(k,p) &=& -a_{6} \frac{ k^{2} + p^{2} }{ (k^{2}-p^{2})^{2} } \left[ A(k^{2}) - A(p^{2}) \right] \, ,   \\ [6pt]
\tau_{7}(k,p) &=& - \left  [ \frac{a_4q^2}{2k^2p^2}  + \frac{a_{7}}{ k^{2}+p^{2} }  \right] \frac{ B(k^{2}) - B(p^{2}) }{ k^{2}-p^{2} } \, , \\ [6pt]
\tau_{8}(k,p) &=& a_{8} \frac{ A(k^{2}) - A(p^{2}) }{ k^{2}-p^{2} } \, ,
\label{Rociotransverse8}
\end{eqnarray}
where the coefficients $a_i$  are unknown constants.

Inserting the above form factors  in Eqs.~\eqref{tau1QCD} to \eqref{tau8QCD}, we derive the following expressions for the $Y_i(k,p)$ which
consequently depend on the $a_i$ coefficients:
\begin{eqnarray}
Y_{1}(k,p) &=& -2 a_{1} \left[ B(k^{2}) - B(p^{2}) \right] \frac{ \nabla (k,p) }{ k^{2} + p^{2} } \  ,
\label{Y-QCD1}  \\ [10pt]
Y_{2}(k,p) &=& \tfrac{1}{2} \left[ A(k^{2}) - A(p^{2}) \right]  \big\{ (k^{2}-p^{2}) \left (G(q^2)X_0(q^2)-2 a_{3} \right )    \nonumber \\ [10pt]
      & - & 2 \left( \frac{ k^{2} + p^{2} }{ k^{2} - p^{2}} \right) (k+p)^{2} a_{6} \Big \} \ , 
\end{eqnarray}      
\begin{eqnarray}
Y_{3}(k,p) &=& \tfrac{1}{2} \left[ A(k^{2}) - A(p^{2}) \right]   \big\{ -q^{2}\left (G(q^2)X_0(q^2)-2 a_{3} \right ) \nonumber  \\ [8pt]
                 &  +& 4 \, \frac{ \nabla (k,p) }{ k^{2} + p^{2} } a_{2} + 2 (k^{2} + p^{2}) a_{6} \big\} \ , \\ [8pt]
Y_{4}(k,p) &=& - \frac{ B(k^{2}) - B(p^{2}) }{ 4 k^{2} p^{2} (k^{2} + p^{2}) }   \big\{ 2 (k^{2} + p^{2}) \nabla (k,p) a_{4} \nonumber \\  [8pt]
                 & + &  2 k^{2} p^{2} (k^{2} + p^{2}) \left [ a_{5} -G(q^2)X_0\right ] +   k^{2} p^{2} (k+p)^{2} a_{7} \big\} \ ,   \\ [8pt]
Y_{5}(k,p) &=& \tfrac{3}{2} \left[ A(k^{2}) - A(p^{2}) \right]   \big\{ -(k-p)^{2} \left [  G(q^2)X_0- 2 a_{3} \right ] \nonumber \\  [8pt]
                        &+&  \frac{4}{3} \frac{ \nabla (k,p) }{ k^{2} + p^{2} } a_{2} + 2 (k^{2} + p^{2}) a_{6} \big \} \ , \\ [8pt]
Y_{6}^{A}(k,p) &=& \frac{  B(k^{2}) - B(p^{2})  }{ 2 k^{2} p^{2} (k^{2} + p^{2}) }  \big\{ 2 (k^{2} + p^{2}) \nabla (k,p) a_{4} \nonumber \\  [8pt]
                       & + & 6 k^{2} p^{2} (k^{2} + p^{2}) \left (a_{5} - G(q^2) X_0  \right ) +   k^{2} p^{2} (k+p)^{2} a_{7} \big \} \ , \\  [8pt]
Y_{7}^{S}(k,p) &=& a_{7} \left[ B(k^{2}) - B(p^{2}) \right] \frac{ k^{2} -p^{2} }{ k^{2} + p^{2} }  \  , \\  [8pt]
Y_{8}^{A}(k,p) & = & - \tfrac{1}{2}\! \left[ A(k^{2}) - A(p^{2}) \right] \! \left (a_{8} + G(q^2) X_0  \right ) \ . 
\label{Y-QCD8}
\end{eqnarray}

While the ansatz given by Eqs.~\eqref{Rociotransverse1} to \eqref{Rociotransverse8} allows us to derive analytic expressions for the $Y_i(k,p)$ form factors,
we trade our ignorance of their functional form---or of the nonlocal tensor $V_{\mu\nu}$ in general---for that of relative coefficients. The task is now to find a set
of solutions $\vec a :=\{a_1, a_2, a_3, a_4, a_5, a_6, a_7, a_8 \}$ that reproduces mass and wave-function renormalization solutions from lattice QCD or
successful  phenomenological models employed in calculations of  hadron properties. It turns out that this set is rather small since some of the $a_i$
are tightly constrained due to multiplicative renormalizability, as we established with the two relations~\cite{Albino:2018ncl},
\begin{eqnarray}
      a_{2} + 2a_{3} -2a_{8}  & = &    0  \ ,
  \label{coeficientconstraint1}        \\
      a_{6}  & = &  \frac{1}{2} \ .
\label{coeficientconstraint2}
\end{eqnarray}
As we shall see, DSE solutions are found in the parameter space defined by the intervals, $-2 \leq a_i \leq +2$.
Allowing for the parameters to leave this range mostly leads to unstable or non-converging iterations when solving the integral equations.


\section{Gauge sector: gluon and ghost propagators from lattice QCD}
\label{ghostgluonsec}

We focus on the complete tensor structures of the dressed quark-gluon vertex, in particular on its contribution to DCSB in the gap equation, and, as such, we do
not attempt to solve the coupled DSEs for the gluon and the ghost. This exercise is certainly most worthwhile and of particular interest to understand the effect of
the vertex in the dressed gluon propagator and its feedback on the quark gap equation. We postpone this more ample and complex study and rather employ the
dressed gluon and ghost propagators which have been calculated on the lattice and for which several numerical results are readily available.

We use the following three data sets for the gluon and ghost propagators in Landau gauge obtained with lattice-QCD simulations:
\begin{description}[leftmargin=!,labelwidth=\widthof{III.}]
\item[I.]  Quenched lattice data by Bogolubsky \emph{et al\/}.~\cite{Bogolubsky:2009dc} generated with $\beta=6 / g_{0}^{2}=5.70$, a lattice spacing $a = 0.17\,$fm
          and a lattice volume of $L^4 = (13.6\,\mathrm{fm})^4$.
\item[II.]  Quenched lattice data for the gluon by Dudal~\emph{et al\/}.~\cite{Dudal:2018cli} generated with $\beta= 6.0$, lattice spacing $a = 0.1016(25)\,$fm and
          two different lattice volumes: $L^4 = (6.57\,\mathrm{fm})^4$  and $L^4 = (8.21\,\mathrm{fm})^4$. Quenched lattice data for the ghost by Duarte
          \emph{et al\/}.~\cite{Duarte:2016iko} generated with $\beta= 6.0$, lattice spacing $a = 0.1016(25)\,$fm and $L^4 = (8.128\,\mathrm{fm})^4$.
\item[III.]  Partially unquenched lattice data by Ayala \emph{et al\/}.~\cite{Ayala:2012pb} from the gauge configurations generated by the European Twisted
          Mass Collaboration (ETMC)
          for $N_f = 2+1+1$ with $\beta= 1.90$, $(L/a)^3 \times T/a = 32^3\times 64$ and  $\beta= 1.95$, $(L/a)^3 \times T/a = 48^3\times 96$.
\end{description}
Each of these three data sets features distinct advantages. The propagators of the sets I and III were calculated with a large lattice volume and the lowest accessible
momenta are $p\approx 70\,$MeV and $p\approx 100\,$MeV, while the largest momenta are $p\approx 4\,$GeV and $p\approx 4.5\,$GeV,  respectively. Data set~III
additionally allows us to analyze unquenching effects in the solutions of the quark DSE. The lattice data of set~II  were obtained with a smaller lattice volume, 
however for a considerably smaller lattice spacing $a$. Therefore, larger momenta up to $p \approx 7.7\,$GeV are accessible, which
provides additional information on the functional behavior in the perturbative domain of these functions.

In exploiting the available lattice data we resort to analytic fits motivated by theoretical considerations. The lattice propagators, $\Delta (q^2)$ and $G(q^2)$, of set
I and II  are parametrized  with an expression that combines the refined Gribov-Zwanziger tree-level gluon propagator in the infrared domain supplemented by
the 1-loop renormalization group behavior at larger momenta. This amounts to a renormalization-group improved Pad\'e approximation and can be written as~\cite{Dudal:2018cli}
\begin{equation}
   \Delta (q^2 )  =  Z \, \frac{q^{2}+M_{1}^{2}}{q^{4}+M_{2}^{2} q^{2}+M_{3}^{4}}
   \left [1 +  \omega \ln \left(\frac{q^{2}+m_\mathrm{gl}^2 }{\Lambda_{\mathrm{QCD}}^2} \right ) \right ]^{\gamma_\mathrm{gl}} \ ,
   \label{gluonfit}
\end{equation}
where $\omega=11 N_c \,\alpha_{s}(\mu) /12 \pi$, $\Lambda_{\mathrm{QCD}}=0.425$~GeV and $\gamma_\mathrm{gl} =-13/22$ is the 1-loop anomalous
gluon dimension. We choose $\mu =4.3$~GeV  and  $\alpha_{s} = 0.3$~\cite{Aguilar:2010cn,Aguilar:2010gm}.
A least-squares fit yields $\chi^2/\mathrm{d.o.f.} =  1.788$ for the renormalized gluon-dressing function of set I~\cite{Bogolubsky:2009dc}:
\begin{align}
    Z               & = 1.440 \pm 0.003 \ ,  &
    M_{1}^{2}  & = 2.880 \pm 0.054 \, \mathrm{GeV}^{2} \ ,  \nonumber \\
    M_{2}^{2}  & = 0.434 \pm 0.021 \, \mathrm{GeV}^{2} \ , &
    M_{3}^{4}  & =  0.527 \pm 0.009 \, \mathrm{GeV}^{4} \ ,  \nonumber \\
    m_\mathrm{gl}^2  & =  0.305 \pm 0.071 \,  \mathrm{GeV}^{2} \  , &
\end{align}
and in case of set II~\cite{Dudal:2018cli} we obtain $\chi^2/\mathrm{d.o.f.} =  0.031$  in a best fit to the \emph{bare\/} gluon propagator with the parameters,
\begin{align}
    Z               & = 8.421 \pm 0.040 \ ,  &
    M_{1}^{2}  & =  2.743  \pm  0.184 \, \mathrm{GeV}^{2} \ ,  \nonumber \\
    M_{2}^{2}  & = 0.519 \pm 0.073 \, \mathrm{GeV}^{2} \ , &
    M_{3}^{4}  & = 0.356 \pm 0.029 \, \mathrm{GeV}^{4} \ ,  \nonumber \\
     m_\mathrm{gl}^2  & =  0.259 \pm 0.262 \,  \mathrm{GeV}^{2} \  , &
\end{align}
where $\Delta (q^2 )$ must be renormalized by a factor $0.3143\, \mu^2$.
Similarly, we make use of the ghost-dressing parametrization introduced in Ref.~\cite{Duarte:2016iko},
\begin{equation}
    G  (q^2)   =   Z\, \frac{q^4+M_2^2 q^2+M_1^4}{q^4+M_4^2 q^2+M_3^4}
     \left [ 1 +\omega \ln \left(\frac{q^2+\frac{m_1^4}{q^2+m_0^2}}{\Lambda_\mathrm{QCD}^2} \right )  \right ]^{\gamma_{\mathrm{gh} } } \  ,
\end{equation}
with the anomalous ghost dimension $\gamma_{\mathrm{gh}} = -9/44$, while $\omega$ and  $\Lambda_\mathrm{QCD}$ take on the same values as in
Eq.~\ref{gluonfit}. Fitting this expression to the ghost propagator of set I~\cite{Bogolubsky:2009dc} renormalized at $\mu =4.3$~GeV  we find
$\chi^2/\mathrm{d.o.f.} = 1.534$ and the following parameters:
\begin{align}
    Z  & =  1.059 \pm 0.001 \ ,  \quad
    M_1^4  =  35.562 \pm 0.464  \,\mathrm{GeV}^4 \ , \quad
    M_2^2  =  33.471 \pm 0.391  \,\mathrm{GeV}^2 \  , \nonumber  \\
    M_3^4  &  =  14.790 \pm 0.188 \,\mathrm{GeV}^4 \ ,  \quad
    M_4^2  =  29.193 \pm 0.331 \, \mathrm{GeV}^2 \ ,  \quad
    m^2_0  =  0.018  \pm  0.010  \,\mathrm{GeV}^2  \ ,  \nonumber \\
    m^4_1  & =  0.001  \pm 0.0002\,\mathrm{GeV}^4 \,  .
\end{align}
Likewise, a least-squares fit to the \emph{bare} ghost propagator of Ref.~\cite{Duarte:2016iko}  yields $\chi^2/\mathrm{d.o.f.} = 0.247$ with
the parameter set:
\begin{align}
    Z  & =   5.068 \pm 0.086 \ ,  \quad
    M_1^4  =  19.281 \pm 3.685  \,\mathrm{GeV}^4 \ , \quad
    M_2^2  =   27.721 \pm 3.388  \,\mathrm{GeV}^2  \ , \nonumber  \\
    M_3^4  &  =  7.695  \pm 1.925 \,\mathrm{GeV}^4 \ ,  \quad
    M_4^2  =  24.340 \pm 2.833 \, \mathrm{GeV}^2 \ ,  \quad
    m^2_0  =   0.527  \pm  0.083  \,\mathrm{GeV}^2 \  ,  \nonumber \\
    m^4_1  & =   0.018  \pm 0.021 \,\mathrm{GeV}^4 \,  .
\end{align}
Accordingly, $G (q^2 )$ must be renormalized by a factor $4.706$ at $\mu = 4.3$~GeV.

On the other hand, we fit the unquenched data in Ref.~\cite{Ayala:2012pb} with a Pad\'e approximant in the region $q \leq q_0 =3\,$GeV and a
renormalization-group improved continuation for  $q > q_0 =3\,$GeV given by
\begin{align}
   \Delta (q^2 ) &  = \,  \frac{ \alpha_1 + \alpha_2 q^ 2}{ \alpha_3 + \alpha_4 q^2 + \alpha_5 q^4} \ ,   &  q \leq q_0 \ , \\
   \Delta (q^2 ) &  = \,  \frac{ \alpha_1 + \alpha_2 q_0^ 2}{ \alpha_3 +  \alpha_4 q_0^2 + \alpha_5 q_0^4} \, \frac{q_0^2}{q^2}
   \left (
   \frac{\log \left (q_0 / \Lambda_\mathrm{QCD} \right )}{\log \left (q/ \Lambda_\mathrm{QCD} \right ) }
   \right )^{\!\!\! \frac{\gamma_{\mathrm{gl}}}{\beta} } \ ,                                   &  q >  q_0 \ ,
\end{align}
where the anomalous dimension including flavor dependence is in this case $\gamma_{\mathrm{gl}} =  \frac{13}{2} - \frac{2}{3} N_f$,
$\beta = 11 - \frac{2}{3} N_f$, $N_f=4$ and $\Lambda_\mathrm{QCD} = 600$~MeV. Analogously, the ghost propagator is fitted to the expressions,
\begin{align}
   G (q^2 ) &  = \,  \frac{ \beta_1 + \beta_2 q^ 2}{\beta_3 + \beta_4 q^2 } \ ,   &  q \leq q_0 \ , \\
   G (q^2 ) &  = \,  \frac{ \beta_1 + \beta_2 q_0^ 2}{\beta_3 + \beta_4 q_0^2 }
     \left (
   \frac{\log \left (q_0 / \Lambda_\mathrm{QCD} \right )}{\log \left (q / \Lambda_\mathrm{QCD} \right ) }
     \right )^{\!\!\! \frac{\gamma_{\mathrm{gh}}}{\beta} } \ ,                                   &  q >  q_0 \ ,
\end{align}
where $\gamma_\mathrm{gh} = 9/4$. The fitted parameters are~\cite{pepefit}:
\begin{align}
    \alpha_1 & = 4.305\; \mathrm{GeV}^2  , \ \    \alpha_2  = 0.979  , \ \  \alpha_3 = 1\; \mathrm{GeV}^4  ,\ \   \alpha_4 =1.581\; \mathrm{GeV}^2  , \ \   \alpha_5 = 1.098 ,  \\
    \beta_1 &= 2.750\; \mathrm{GeV}^2 , \  \   \beta_2  =1.855 \ ,  \ \ \beta_3 = 1\; \mathrm{GeV}^2 ,  \ \  \beta_4 = 1.864  \ .
\end{align}


\section{Results}
\label{sec4}

\subsection{Solving the Dyson-Schwinger equation for real space-like momenta }

The DSE~\eqref{finalDSEquark}  is solved by first projecting out the scalar functions $A(p^2)$ and $B(p^2)$ of the general solution~\eqref{DEsol}. This yields two
coupled, nonlinear integral equations which additionally couple to the integral equation for $X_0(q^2)$~\eqref{X0int}. We solve these three equations,
$\mathcal{F}(p^2) = A(p^2), B(p^2)$ and $X_0(q^2)$,  simultaneously via an iterative procedure using as convergence criterium,
\begin{equation}
    \bm  \epsilon_\mathcal{F} =  \dfrac{\left \vert \mathcal{F}^{n+1} (p^2 ) - \mathcal{F}^n (p^2 ) \right \vert }{\mathcal{F}^n (p^2 )}  = 10^{-3} \ ,
\end{equation}
where $n=20-30$ iterations suffice. Imposing the constraints~\eqref{coeficientconstraint1} and \eqref{coeficientconstraint2} we have established the following
values for the coefficients $a_i$:
\begin{equation}
\centering
\bgroup\renewcommand*{\arraystretch}{1.5}\renewcommand*{\arraycolsep}{8pt}
\begin{array}{c|c|c|c|c|c|c|c}    \hline  \hline
   a_1 & a_2 & a_3 & a_4 & a_5 & a_6 &  a_7 & a_8  \\ \hline
   2   & 2  & -1 & -1 & 1 & \tfrac{1}{2} & \tfrac{1}{2} & 0  \\ \hline  \hline
\end{array}
\egroup
\end{equation}
The crucial choice is $a_3 \leq 0$, as positive values of this coefficient inevitably lead to spurious or non-converging solutions of the integral-equation system.
With the choice $a_3=-1$ and with $a_6$ fixed by the condition~\eqref{coeficientconstraint2} we then adjust $a_2$ and $a_8$ to satisfy condition~\eqref{coeficientconstraint1}.
It turns out that the wave-function renormalization, $Z(p^2)$, is quite sensitive to $a_8 \neq 0$ and either positive or negative values impact its functional behavior
unfavorably in the infrared domain; i.e. $Z(p^2)$ turns back to its perturbative value $\approx 1$, a feature also observed in conjunction with the Ball-Chiu vertex.
A similar observation holds for $a_4 >0$ and we therefore use $a_4 = -1$. Thus, setting $a_8=0$ we have $a_2 = 2$ while $a_1>0$ leads to an increase in the
mass function.  The solutions exhibit little dependence on the choice for $a_5$, whereas $a_7 <0$ results in a mass function that decreases too fast and yields
negative masses at large momenta. We remark that this choice is not unique, but within the constraints of Eqs.~\eqref{coeficientconstraint2} and
\eqref{coeficientconstraint2} and with the necessary choice of $a_3 \leq 0$ little room for variation is left as we checked with some $n> 200$ configurations.

\begin{figure}[t!]
\vspace*{-1cm}
\centering
  \includegraphics[scale=0.86]{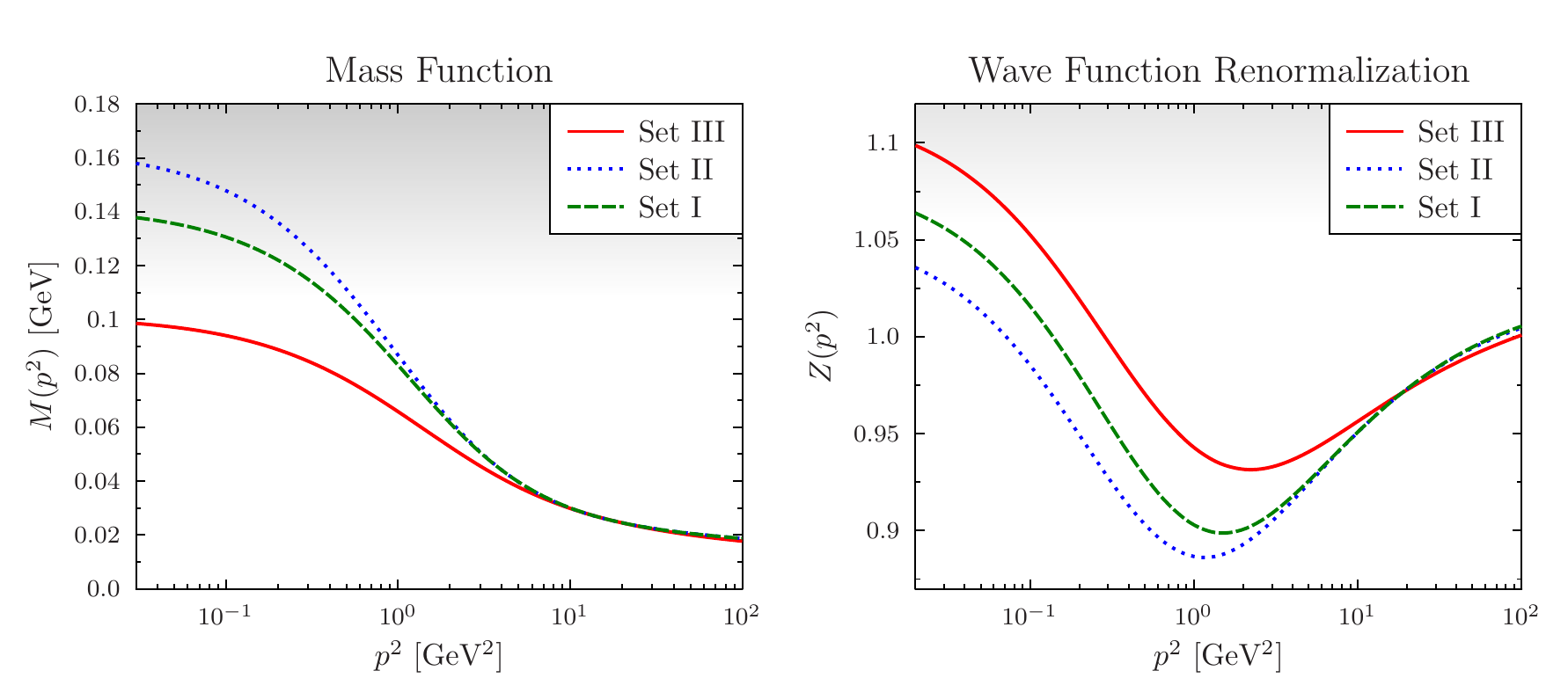}  \vspace*{-2.2cm} \\
  \includegraphics[scale=0.86]{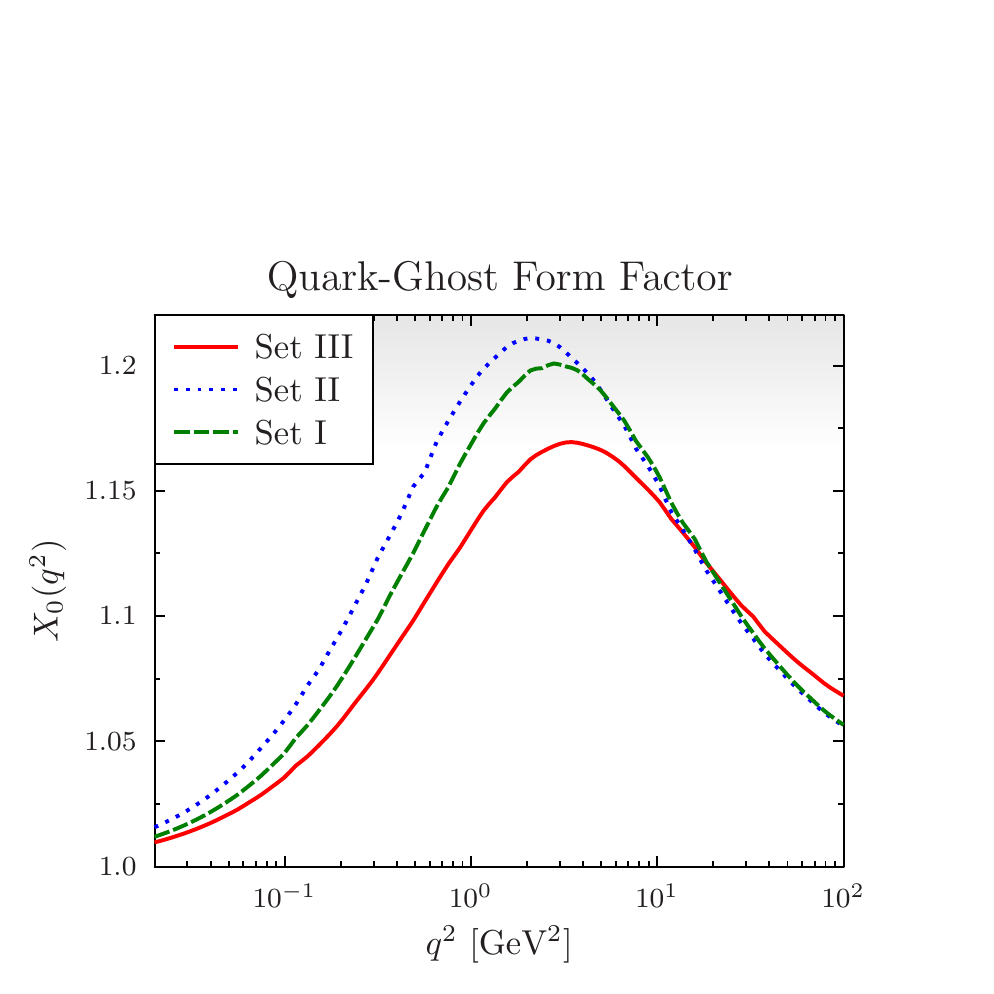}  
  \caption{Mass function, $M(p^2)$, wave renormalization, $Z (p^2)$, and the quark-ghost kernel form factor, $X_0(q^2)$, for three sets of gluon
               and ghost propagators, using only $\Gamma_\mu^L (k,p)$ and the form factors of Eqs.~\eqref{lambda1QCD}--\eqref{lambda4QCD}
               in the DSE~\eqref{finalDSEquark}.
               Set I~\cite{Bogolubsky:2009dc} and II~\cite{Dudal:2018cli,Duarte:2016iko}: $\alpha_s(\mu) = 0.30$,  $m(\mu) = 25\,$MeV, $\mu =4.3$\, GeV.
               Set III~\cite{Ayala:2012pb}: $\alpha_s(\mu) = 0.35$,  $m(\mu) = 25\,$MeV, $\mu =4.3$\, GeV.    }
\label{fig1}
\end{figure}

Our quark-gluon vertex is now fully determined and in the following we present the solutions for $M(p^2)$, $Z(p^2)$ and $X_0(q^2)$ for the three sets of gluon
and ghost propagators discussed in Section~\ref{ghostgluonsec}. Note that in what follows we determine $Z_4(\mu, \Lambda)$ by imposing $m(\mu) \equiv M(\mu)$
at $\mu=4.3$~GeV. Considering degenerate light-quark masses in the range $m(2\,\mathrm{GeV})  \approx $ 30--50~MeV employed in lattice simulations~\cite{Ayala:2012pb},
we choose $m(4.3\,\mathrm{GeV}) =25$~MeV. Along with $A(4.3\,\mathrm{GeV}) = 0.98$, we therefore impose the renormalization conditions:
\begin{eqnarray}
  Z_2 (\mu,\Lambda ) &  = &  \frac{A(\mu^2 )}{1 + \Sigma_{\rm v} (\mu,\Lambda ) } \ ,
  \label{Z2renorm}   \\
  Z_4 (\mu,\Lambda )  &  = &  \frac{B(\mu^2 ) -   Z_2(\mu,\Lambda )\,  \Sigma_{\rm s} (\mu,\Lambda ) }{m(\mu)} \ ,
\label{Z4renorm}
\end{eqnarray}
where $\Sigma_{\rm s} (\mu,\Lambda )$  and $\Sigma_{\rm v} (\mu,\Lambda )$ are the integrals after the usual scalar and vector projections of the right-hand-side of
Eq.~\eqref{finalDSEquark} with $\mathcal{P}_B = \mathbbm{1}_D/4$ and $ \mathcal{P}_A = -i\gamma\cdot p/4p^2$, respectively:
\begin{align}
\label{A-sigma}
   A ( p ^2 ) \ &  = \ Z_2(\mu,\Lambda )   +  Z_2(\mu,\Lambda ) \, \Sigma_{\rm v}  ( p ,\Lambda  ) \  ,  \\
\label{B-sigma}
   B ( p ^2 ) \ & =  \ Z _4(\mu,\Lambda ) \, m( \mu)  +  Z_2(\mu,\Lambda )\, \Sigma_{\rm s} ( p, \Lambda ) \ .
\end{align}

In order to appreciate the effect of the non-transverse components of $\Gamma_\mu(k,p)$, we first  omit the contributions of $\Gamma_\mu^T$ following
the work of Ref.~\cite{Rojas:2013tza}, however without artificially enhancing $X_0(q^2)$ by an inversion procedure. The mass and wave-renormalization functions
along with the form factor $X_0(q^2)$ are plotted in Figure~\ref{fig1} for the gluon and ghost propagator sets discussed in Section~\ref{ghostgluonsec}.
The values for $m(\mu)$ and $\alpha_s(\mu)$ at the renormalization scale $\mu$ are detailed for each set in Figure~\ref{fig1} and we remark the following:
the gluon- and ghost- dressing functions of sets I and II are renormalized at $\mu = 4.3\,$ GeV where we employ $\alpha_s (\mu) = 0.30$ at this scale following
Refs.~\cite{Aguilar:2010cn,Aguilar:2010gm}. We also renormalize the DSE at $\mu =4.3$~GeV when using the unquenched propagators of set III but increase
the strong coupling to $\alpha_s (\mu)= 0.35$, as otherwise only a trivial solution, $M(0) \approx 0$, is found. This is in line with the behavior of the Taylor coupling in 
Ref.~\cite{Ayala:2012pb}.

The mass functions in Figure~\ref{fig1} exhibit the well-known fast rise at about $p=1$~GeV and saturate at low momenta, where the mass $M(0)$ ranges
between 100~MeV and 163~MeV. While these solutions exhibit appreciable DCSB, it is much lower than that obtained with phenomenological models
and does not yield the correct pion mass and weak decay constant~\cite{Bashir:2012fs}. The functional behavior of $Z(p^2)$ is characteristic of that produced
with a Ball-Chiu vertex and the Maris-Tandy model~\cite{Maris:1997tm} in the gap equation and indeed, in Eqs.~\eqref{lambda1QCD} to \eqref{lambda4QCD}
the longitudinal form factors describe a ``ghost-corrected'' Ball-Chiu vertex. However, the depth of the minimum of $Z(p^2)$ is less pronounced than with
the phenomenological approach and the growth in the deep infrared is exacerbated with values of $Z(0) >1$. Clearly, limiting the vertex to its non-transverse
components yields mass- and wave-renormalization functions which are unlike those required for realistic applications in hadron physics. We note that
the leading form factor of the quark-ghost scattering kernel behaves similarly in all three cases with a maximum located around 2--5~GeV$^2$ but different magnitudes.
Notably, the more enhanced solutions for $X_0(q^2)$ using set I and II are also accompanied by an increased DCSB in the corresponding mass functions.
The net contribution of $X_0(q^2)$ to $M(p^2)$ is 27\% when using set I and 29\% when using set II; solving the DSE with set III and keeping $X_0 (q^2)=1$
does not yield a mass gap.

\begin{figure}[t!]
\vspace*{-1.2cm}
\centering
  \includegraphics[scale=0.86]{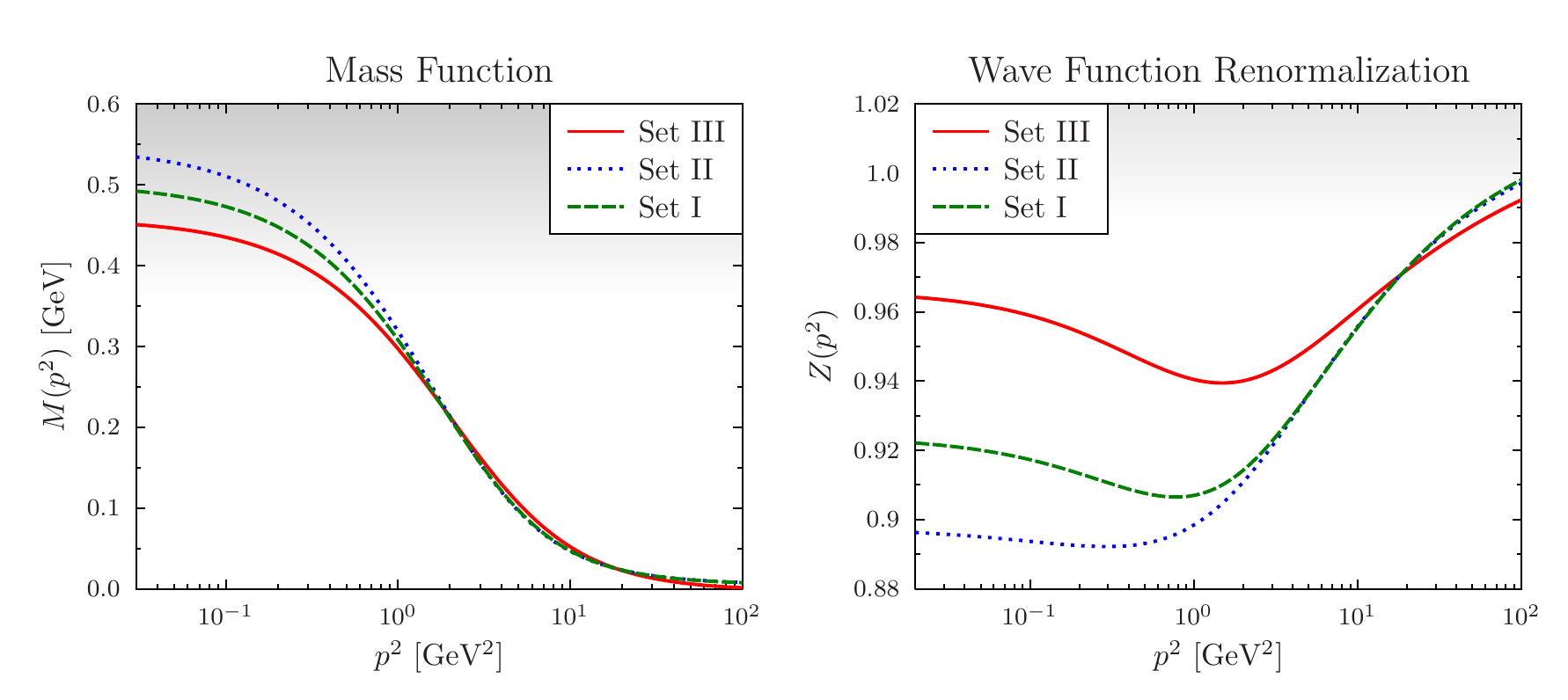}  \vspace*{-2.6cm} \\
  \includegraphics[scale=0.86]{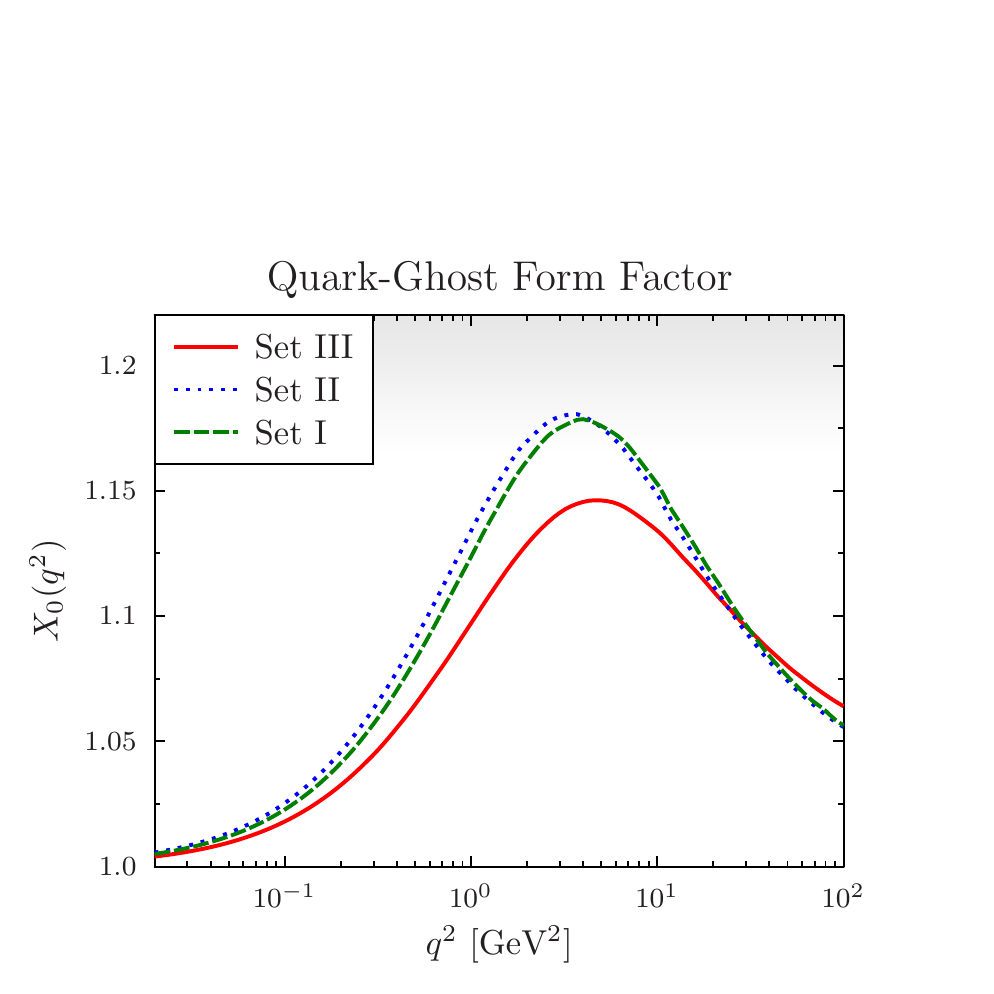}  \vspace*{-3mm}
  \caption{Mass function, $M(p^2)$, wave renormalization, $Z (p^2)$, and leading quark-ghost kernel form factor, $X_0(q^2)$, obtained with three sets of gluon
               and ghost propagators, $\Gamma_\mu(k,p) =  \Gamma^{L}_{\mu}(k,p)  + \Gamma^{T}_{\mu}(k,p)$ and associated form factors of
               Eqs.~\eqref{lambda1QCD}--\eqref{tau8QCD} in the DSE~\eqref{finalDSEquark}. Current-quark
               masses, strong couplings and renormalization scale as in Figure~\ref{fig1}. }
\label{fig2}
\end{figure}

We now introduce the transverse components of the quark-gluon vertex, $\Gamma_\mu(k,p) =  \Gamma^{L}_{\mu}(k,p)  + \Gamma^{T}_{\mu}(k,p)$, in
the DSE and therefore all 12 form factors, Eqs.~\eqref {lambda1QCD} to \eqref{lambda4QCD} and \eqref{tau1QCD} to \eqref{tau8QCD}, come into play.
The solutions for $M(p^2)$, $Z(p^2)$ and $X_0(q^2)$ are again juxtaposed for the three sets of gluon and ghost propagators in Figure~\ref{fig2}.
We immediately notice the drastic increase of the mass functions, exemplified by $M(0) = 500$~MeV for set~I, $M(0) =  546$~MeV for set II and $M(0)=458$~MeV
for set III. In other words, including the transverse vertex leads to a 235\%--350\% increase in DCSB, while the contribution of the quark-ghost scattering amplitude
$X_0(q^2)$ is in the range of 14\%--17\%. The resulting wave renormalizations likewise exhibit a different behavior at low momenta: while they do not monotonously
fall as lattice-QCD calculations indicate, albeit with larger errors at low momenta~\cite{Oliveira:2016muq}, they flatten out and only slightly bend over compared
to the solutions obtained with merely the longitudinal vertex. The quark-ghost form factors $X_0(q^2)$ in Figure~\ref{fig2} are similar to those in Figure~\ref{fig1},
yet we note that overall their maximum values decrease. 

A worthwhile observation is that no solutions for the coupled integral equations~\eqref{A-sigma}, \eqref{B-sigma} and \eqref{X0int} are found if we define 
$Y_i(k,p) \equiv 0$, $i=1,...,8$, regardless of the value  of the strong coupling $\alpha_s(\mu)$. The same is true if we set $\tau_4(k,p) =  \tau_7(k,p) =0$. 
Indeed, these two transverse form factors are responsible for the \emph{overwhelming contribution\/} to DCSB in the gap equation. Keeping them both and discarding, one by one, 
the remaining $\tau_i (k,p)$, $i=1,2,3,5,6,8$ or combinations thereof, results in very similar solutions. Depending on which form factor is discarded, variations of 5--10\% 
of the mass functions plotted in Figure~\ref{fig2} are observed, some leading to a decrease while others to an increase of $M(p^2)$. Hence, the  form factors other than 
$\tau_4(k,p)$  and  $\tau_7(k,p)$ are merely responsible for the fine details of the transverse vertex. We note that $\tau_4(k,p)$ and  $\tau_7(k,p)$ are very similar in structure 
to $Y_4(k,p)$, $Y_6^A(k,p)$ and $Y_7^S(k,p$), all of which are proportional to the difference of the mass functions, $\Delta B(k,p) =B(k) - B(p)$. On the other hand, if we retain only  
$\tau_3(k,p)$  and $\tau_5(k,p)$, which also explicitly depend on $\Delta B(k,p)$, the resulting mass functions exhibit modest DCSB, about 20--30\% larger than when 
the transverse vertex is completely neglected, as depicted in Figure~\ref{fig1}.

\begin{figure}[t!]
\vspace*{-1cm}
\centering
  \includegraphics[scale=0.85]{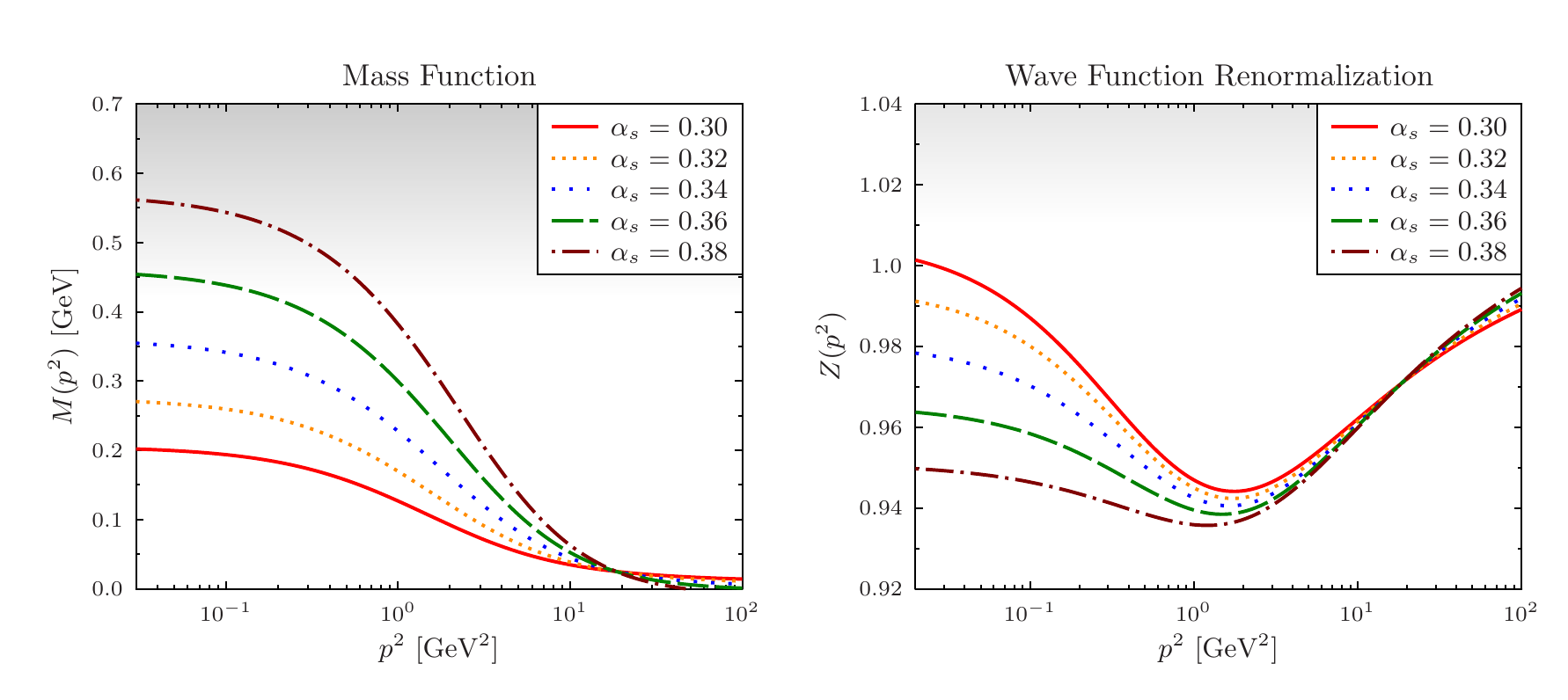}  \vspace*{-2.6cm} \\
  \includegraphics[scale=0.85]{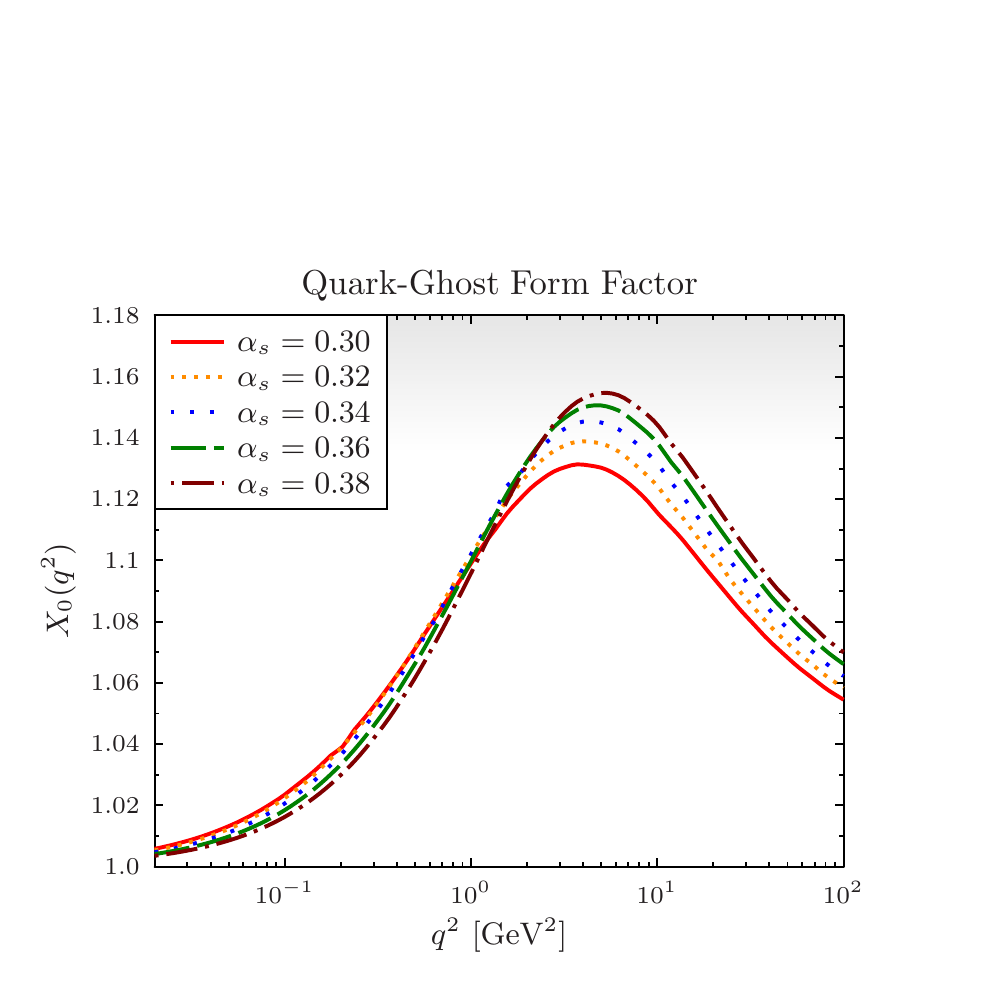}   \vspace*{-3mm}
  \caption{Functional dependence of $M(p^2)$,  $Z (p^2)$ and $X_0(q^2)$ on the strong coupling $\alpha_s$. The DSE~\eqref{finalDSEquark} is solved with the
                vertex  $\Gamma_\mu(k,p) = \Gamma^{L}_{\mu}(k,p)  + \Gamma^{T}_{\mu}(k,p)$ and the corresponding form factors in Eqs.~\eqref{lambda1QCD}--\eqref{tau8QCD}; 
                and with the gluon and ghost propagators of set III~\cite{Ayala:2012pb}. The current-quark mass is
                $m(4.3\,\mathrm{GeV}) = 25$~MeV.}
\label{fig3}
\end{figure}

\begin{figure}[t!]
\vspace*{-1cm}
\centering
  \includegraphics[scale=0.86]{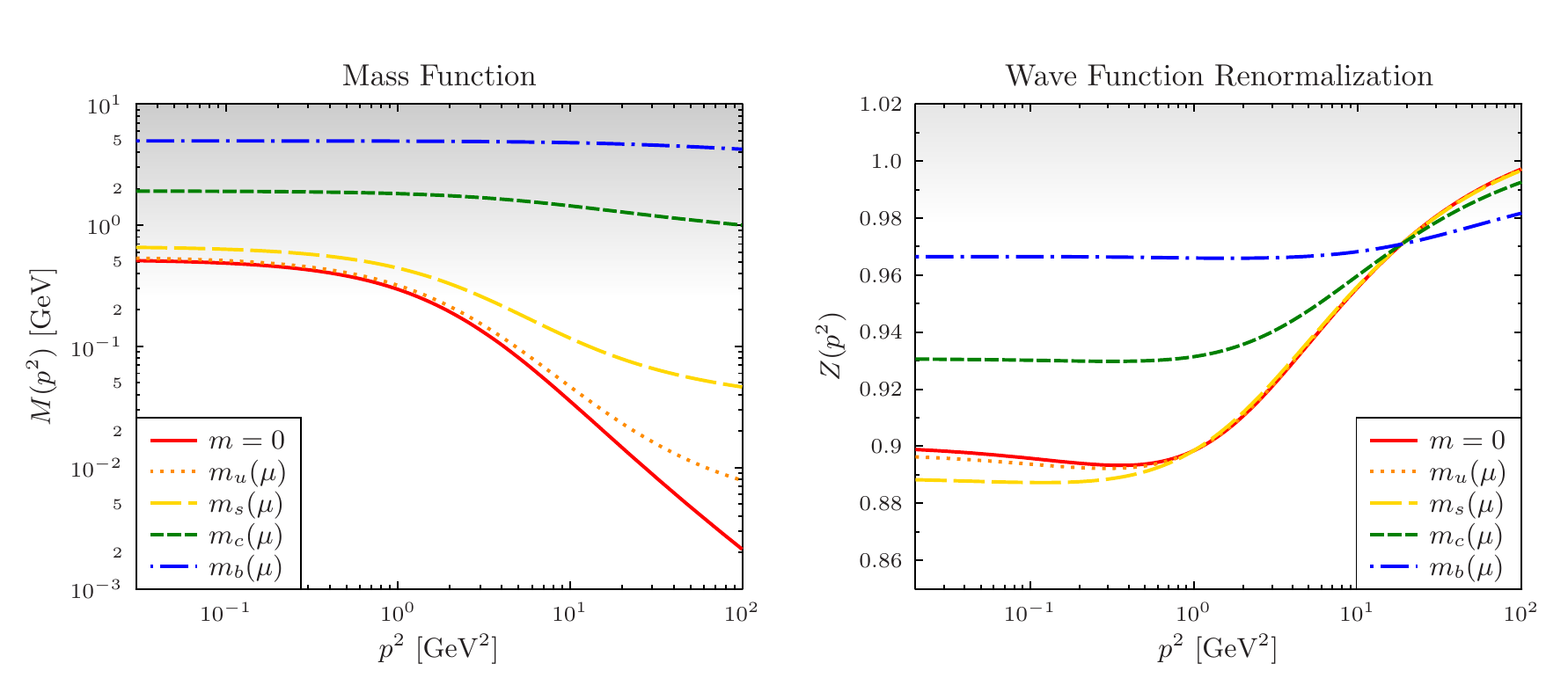}  \vspace*{-2.6cm} \\
  \includegraphics[scale=0.86]{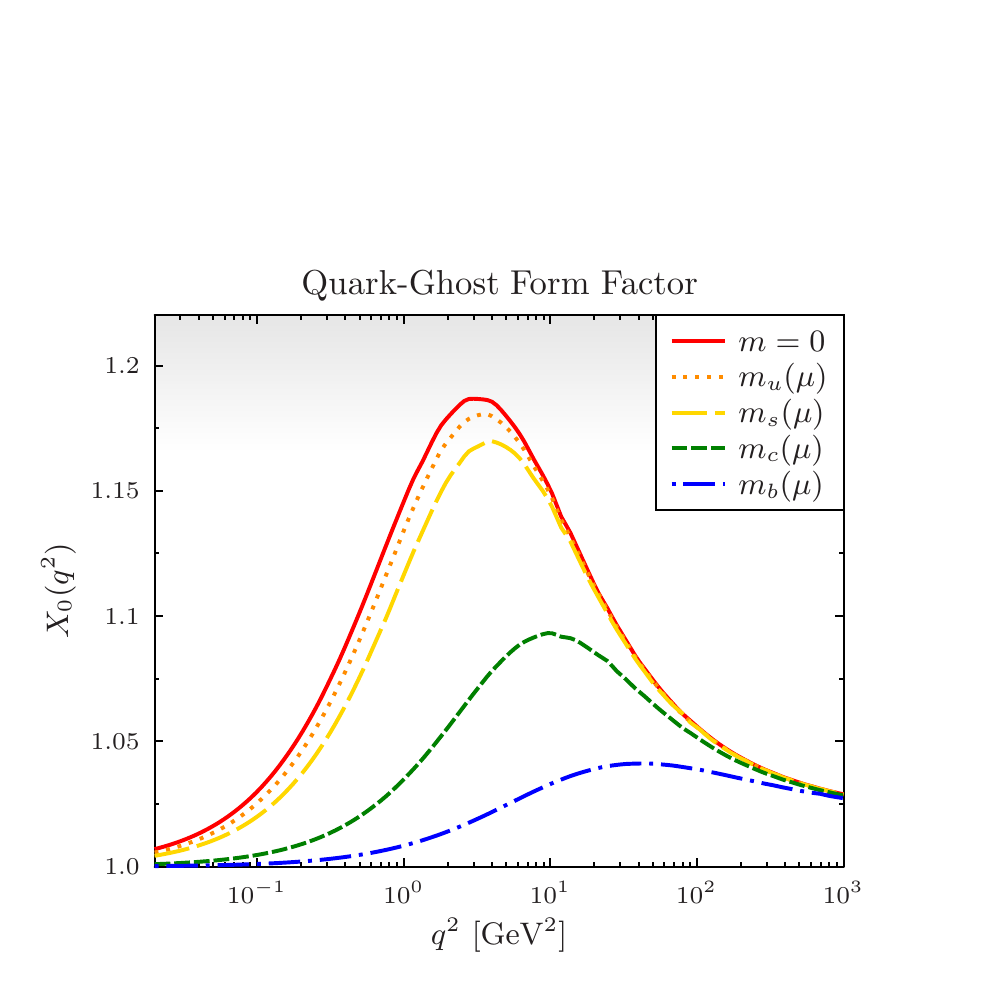}   \vspace*{-3mm}
  \caption{Flavor dependence of $M(p^2)$,  $Z (p^2)$ and $X_0(q^2)$ obtained with the gluon- and ghost propagators of set II~\cite{Dudal:2018cli,Duarte:2016iko}.
               The DSE~\eqref{finalDSEquark} is solved with the vertex $\Gamma_\mu(k,p) = \Gamma^{L}_{\mu}(k,p) + \Gamma^{T}_{\mu}(k,p)$ and the associated form factors in
               Eqs.~\eqref{lambda1QCD}--\eqref{tau8QCD}. The strong coupling is $\alpha_s(\mu) = 0.3$ and the current quark masses are: $m_{u,d}(\mu) = 25.0$~MeV, 
               $m_s(\mu) = 82.1$~MeV, $m_c(\mu) = 1.304$~GeV and $m_b(\mu) = 4.697$~GeV at the renormalization scale $\mu=4.3$~GeV. For comparison, we also plot 
               the solutions in the chiral limit which define the quark propagators in Eqs.~\eqref{fpi} and \eqref{qqcondensate}.  }
\label{fig4}
\end{figure}

The mass functions in Figure~\ref{fig2} clearly demonstrate that the inclusion of the transverse components in the DSE has a dramatic impact on dynamical mass
generation. This was already realized long ago and calculations beyond the rainbow-ladder truncation commonly include phenomenological ans\"atze  for
several transverse components, in particular those proportional to $\Delta B(k^2,p^2) = B(k^2) - B(p^2)$. The novelty in this approach is that the present
transverse quark-gluon vertex, derived from gauge identities and whose unknown tensor component is constrained with a vertex ansatz motivated by perturbation 
theory and multiplicative renormalizability~\cite{Bashir:2011dp}, generates copious DCSB without the need to resort to a phenomenological gluon model.

On the other hand, the wave renormalization $Z(p^2)$ tends to increase again in the infrared region below 1~GeV in case of the gluon and ghost propagators of
set I and III.  One may wonder whether the strength of the transverse vertex, whose effect flattens out $Z(p^2)$ at lower momenta, is compatible with what 
model interactions in given truncation schemes predict. The overall strength of the quark-gluon interaction, however, is controlled by the strong coupling. 
In using what lattice QCD offers us for the gluon and ghost propagators renormalized at 4.3~GeV, we also work with a strong coupling expected at that 
renormalization scale. However, given that our transverse vertex has not been determined \emph{exactly\/}  and that our approach is not self-consistent---we do 
not solve the gluon and ghost DSE that couple to the quark-gap equation---we may consider variations of $\alpha_s$. We do so in Figure~\ref{fig3} where we vary 
the coupling $\alpha_s(4.3\,\mathrm{GeV})$  in the range between 0.30 and 0.38 and concentrate merely on set III of gluon and ghost propagators. Plainly, an increasing
coupling strength carries along two modifications:  $M(p^2)$ is enhanced, which is expected,  whereas $Z(p^2)$ is increasingly suppressed and eventually
levels out below $p\approx 1$~GeV. The form factor $X_0(q^2)$ is also enhanced as a function of $\alpha_s$ and its peak value slightly shifts to larger momenta.

The DSE can obviously also be solved for different quark flavors and we do this exercise in Figure~\ref{fig4}, where we make use of the gluon and ghost
propagators of set II; the results for the two other sets are qualitatively and quantitatively very similar. We employ the light, strange and charm current-quark masses
renormalized at 2~GeV of Ref.~\cite{Ayala:2012pb}: $m_u (2\,\mathrm{GeV}) =m_d (2\,\mathrm{GeV}) = 40$~MeV, $m_s(2\,\mathrm{GeV}) =95$~MeV,
$m_c (2\,\mathrm{GeV}) =1.51$~GeV evolved to the scale $\mu = 4.3$~GeV. For the bottom quark we choose $m_b (4.3\,\mathrm{GeV}) = M_b(4.3\,\mathrm{GeV}) = 4.70$~GeV,
where $M_b(p^2)$ is the mass function for that flavor obtained with an interaction model~\cite{Qin:2011dd} in rainbow truncation.

The pattern of the mass function for increasing current-quark masses mirrors the well-known behavior predicted by phenomenological gluon and vertex models,
see Refs.~\cite{ElBennich:2012tp,Serna:2018dwk} for example. At $p=0$~GeV we find: $M_u(0) =0.55$~GeV, $M_s(0) =0.67$~GeV, $M_c(0) =1.92$~GeV
and $M_b(0) =4.97$~GeV. Thus, even for the charm quark the impact of dynamical mass generation is important. The wave renormalization function flattens
out in the infrared region for heavier quarks but remains above $Z(0) \approx 0.93$ when  $Z(4.3\,\mathrm{GeV})  = 0.98$ is imposed.
We also read from Figure~\ref{fig4} that $X_0(q^2)$ is very sensitive to the current-quark mass and while the function is increasingly damped, its maximum value
is considerably shifted to large momenta where it contributes little to DCSB in heavy quarks.


\subsection{Solving the Dyson-Schwinger equation on the complex plane}


\begin{figure}[t!]
 \vspace*{-1cm}
\centering
 \hspace*{2mm}
  \includegraphics[scale=0.4]{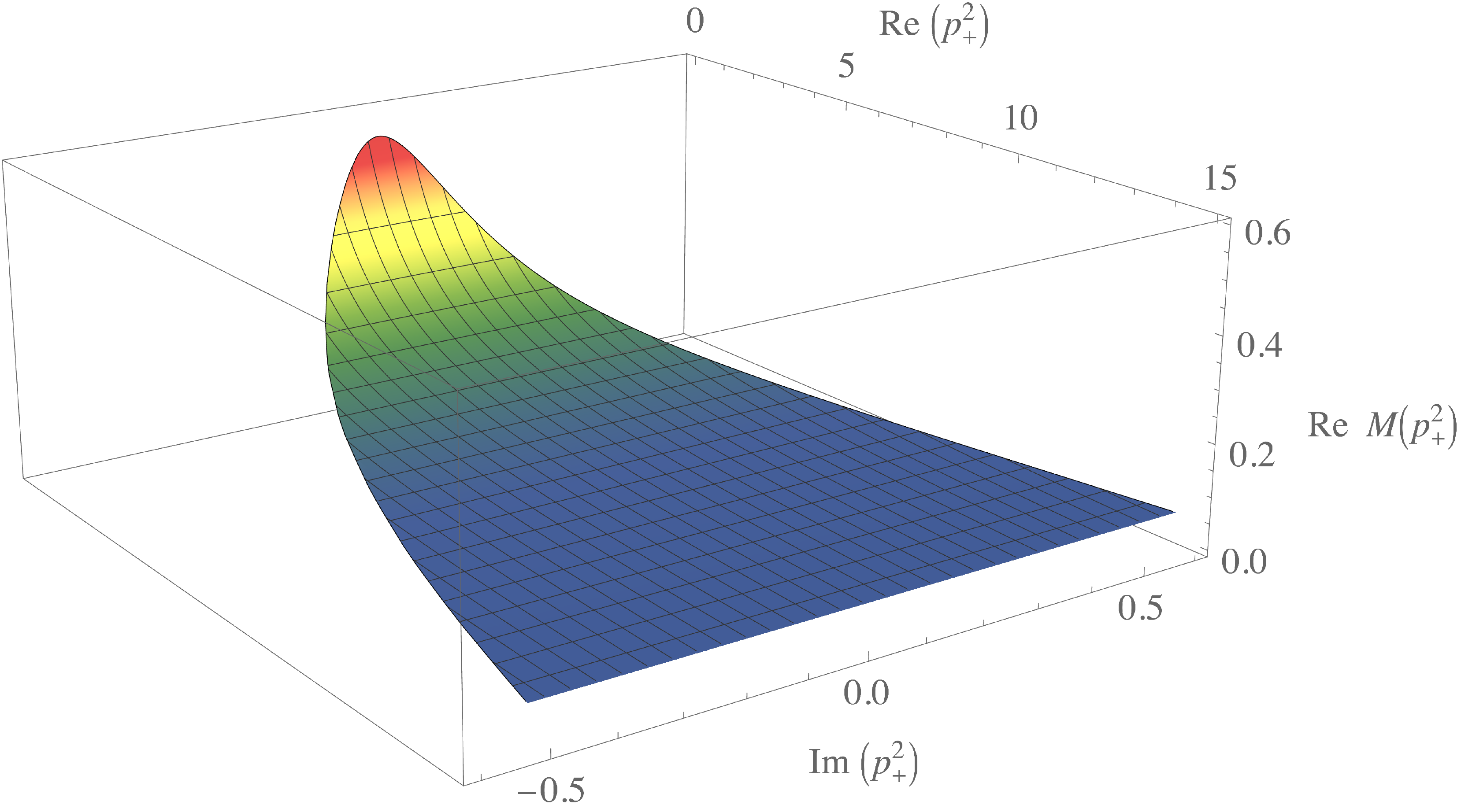}   \vspace*{0.3cm} \\
 \includegraphics[scale=0.4]{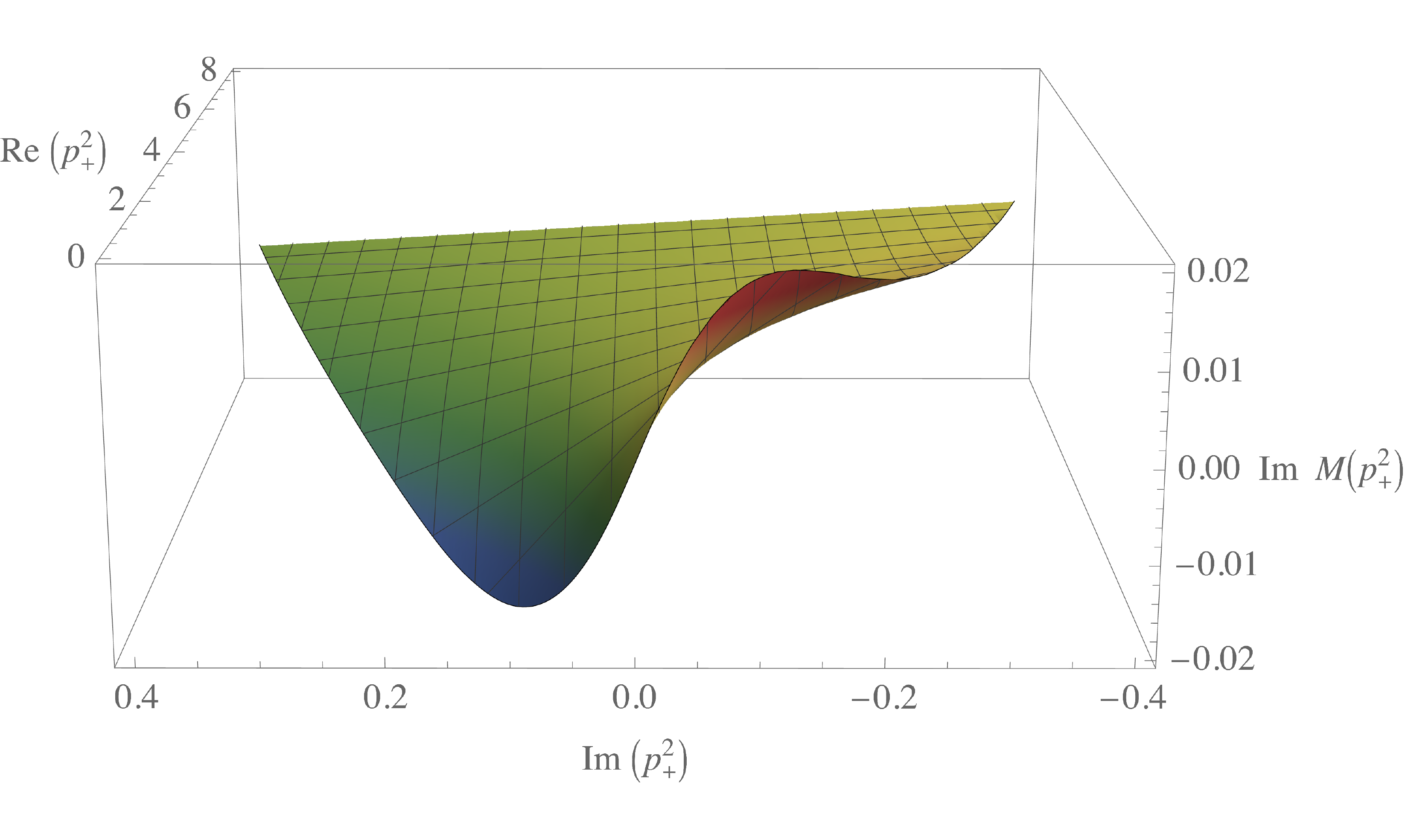}   \vspace*{-0.7cm}
\caption{Real part (upper panel) and imaginary part (lower panel) of the complex mass function $M (p^2)$. The DSE is solved on the complex parabola
with $p_+$ defined  by Eq.~\eqref{BSEmomentum}, the external mass $P^2 = -M_\pi^2$ ($M_\pi = 140$~MeV) and $\eta =1/2$. We use the gluon and ghost 
propagators of set II~\cite{Dudal:2018cli,Duarte:2016iko}, the vertex $\Gamma_\mu(k,p) = \Gamma^{L}_{\mu}(k,p) + \Gamma^{T}_{\mu}(k,p)$ with the form factors in
Eqs.~\eqref{lambda1QCD}--\eqref{lambda4QCD} and \eqref{tau1QCD}--\eqref{tau8QCD}, while $\alpha_s(\mu) = 0.3$. Mass units are in GeV.}
  \label{fig5}
\end{figure}

So far, we have solved the DSE on the real spacelike axis which suffices to evaluate the magnitude of DCSB and the confining properties of the solutions.
However, in solving the bound-state equation in Euclidean space~\cite{Serna:2017nlr} the quark propagators,
\begin{equation}
   S(p_\pm)  =  -i \gamma \cdot p_\pm \, \sigma_\mathrm{v} (p_\pm^2 ) + \sigma_\mathrm{s}  (p_\pm^2 ) \ ,
\end{equation}
are evaluated at complex-valued momenta. This is because in Euclidean space their arguments, $p_+ = p + \eta P$ and $p_- = p - \bar \eta P$, define parabolas
on the complex plane,
\begin{align}
     p^2_+   =   p^2  - \eta^2 M^2  + 2 i \eta  M  | p | z_p \  ,    \label{BSEmomentum}    \\
     p^2_-   =   p^2  - \bar \eta^2 M^2  - 2 i  \bar \eta  M  | p | z_p \  ,
\end{align}
where $P= \big( \vec 0, i M \big )$ is the meson's rest-frame momentum, $\eta + \bar \eta = 1$  are the momentum partition parameters with $\eta, \bar \eta \in [0,1]$,
and $z_p = p \cdot P /|p||P|$ is the angle between the relative and total meson momenta for which $-1 \leq z_p \leq +1$ holds.


\begin{figure}[t!]
 \vspace*{-0.5cm}
 \centering
 \hspace*{4mm} \includegraphics[scale=0.4]{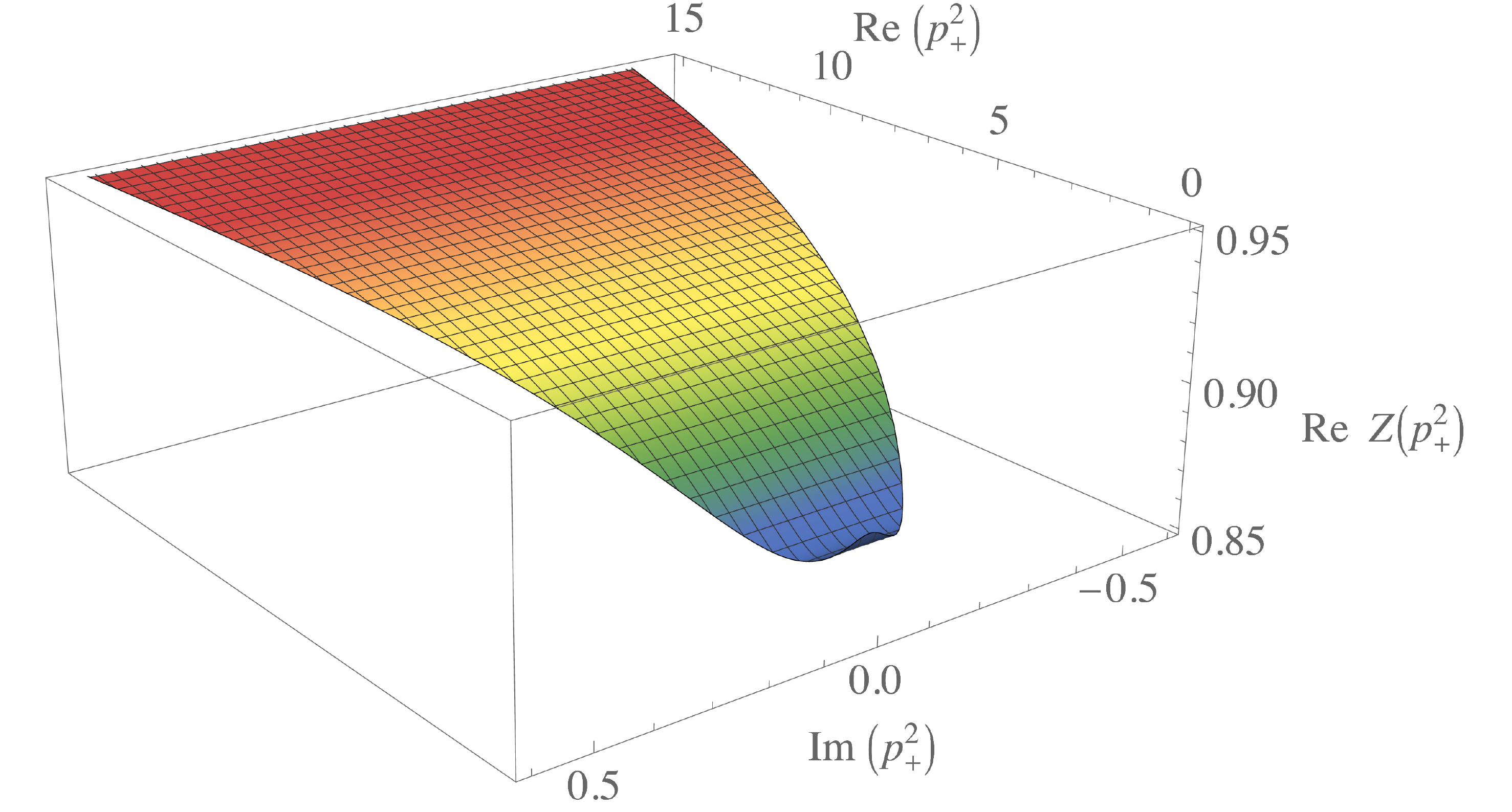}   \vspace*{0.3cm} \\
 \hspace*{-6mm}  \includegraphics[scale=0.4]{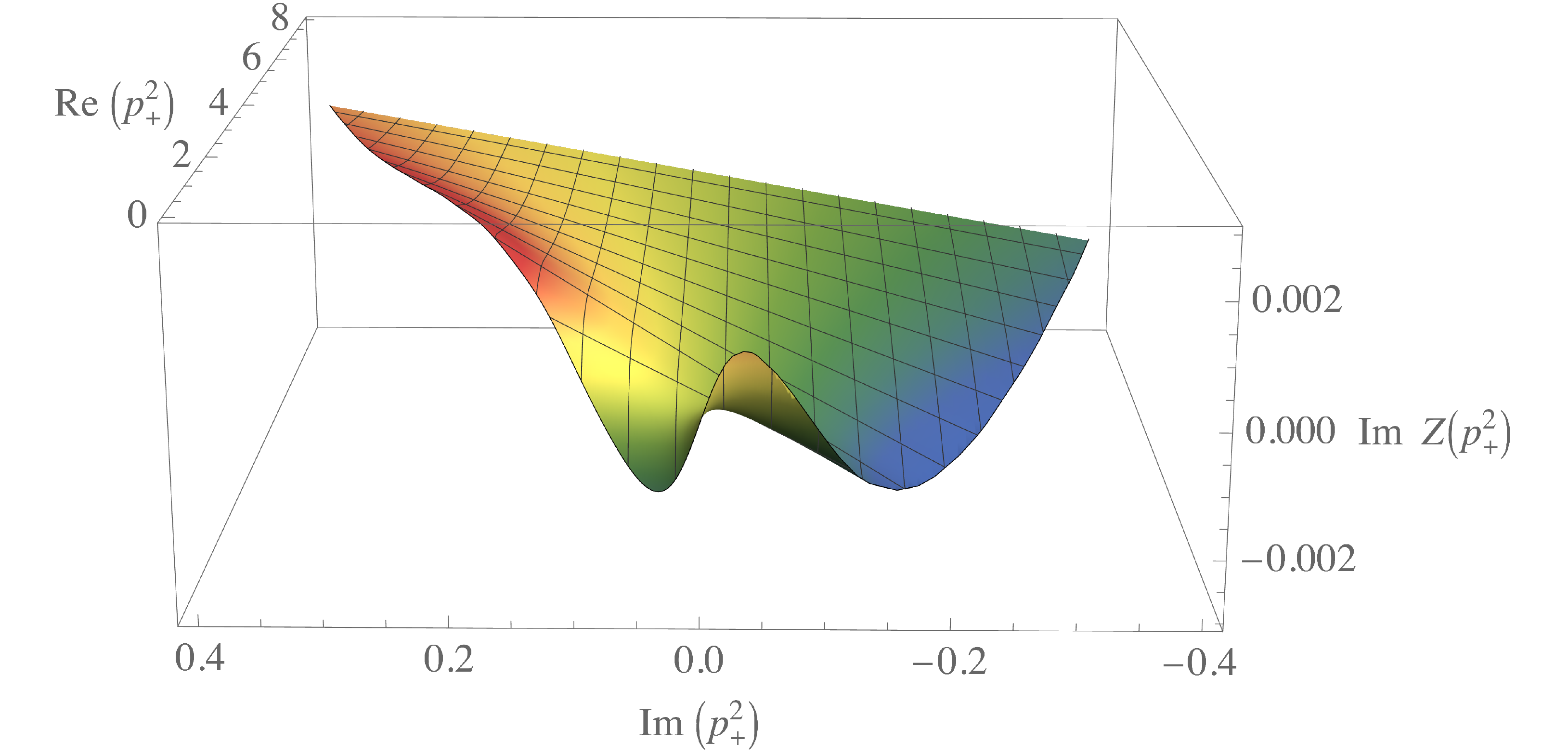}
\caption{Real part (upper panel) and imaginary part (lower panel) of the complex wave-renormalization function $Z(p^2)$. Details of the DSE solution as in Figure~\ref{fig5}.  }
  \label{fig6}
\end{figure}


We use Cauchy's integral theorem to solve the quark DSE inside this parabola. At first, we parametrize the contour of the parabola in the lower and upper
half of the complex plane and find solutions on this contour via an iterative procedure using the same renormalization procedure as in Eqs.~\eqref{Z2renorm}
and \eqref{Z4renorm}. Then, we apply this contour solution in Cauchy's integral formula again to find solutions for  $\sigma_\mathrm{v} (p_\pm^2 )$  and
$\sigma_\mathrm{s}  (p_\pm^2 )$ inside the parabola.  A detailed discussion with examples of parametrizations can be found in Refs.~\cite{Fischer:2005en,
Krassnigg:2009gd,Rojas:2014aka}.

The real and imaginary parts of the $M(p^2)$ and $Z(p^2)$ functions are plotted in Figures~\ref{fig5} and \ref{fig6}, respectively. The real part  of $M (p^2)$ is a
smooth and monotonically decreasing function of $p^2 \equiv p^2_+$ in real direction, whereas it is all but constant along the imaginary axis $\mathrm{Im}\, (p^2)$.
Likewise, the real part of $Z(p^2)$ smoothly tends towards its perturbative limit both on and off the real axis. The imaginary part of both functions is characterized
by complex-conjugate extrema near the origin of the parabola, though we note that their magnitude is considerably smaller than that of the imaginary parts of solutions
with model propagators~\cite{Maris:1997hd,Qin:2011dd,El-Bennich:2020aiq}.

Generally speaking, in the complex momentum range considered herein, the two solutions presented in Figures~\ref{fig5} and \ref{fig6} are qualitatively very similar
to those obtained with phenomenological interaction models in rainbow-ladder truncation. There are, on the other hand, distinctive quantitative and analytical differences
and we here refrain from a detailed comparison. We merely note that the convergence of the complex DSE using a Cauchy integral, with the present vertex and its inherent 
complexities, raises subtle issues that depend on the details of the contour parametrization and its size determined by the parabola cutoff on the complex plane.


\begin{table}[b!]
\centering
\begin{tabular}{R{2.7cm}|C{1.3cm}|C{1.3cm}|C{1.3cm} } \hline 
                   &  Set I  & Set II & Set III \\ \hline\hline
   $f_\pi^0$  [MeV]    &  97.0 &  96.02   &   98.67  \\   \hline
   $ (-\langle \bar qq \rangle)^{\frac{1}{3}}$ [MeV]   &  251.16 &  249.61 &  255.80  \\
   \hline
\end{tabular}
\caption{Weak decay constant in the chiral limit~\eqref{fpi} and quark condensate~\eqref{qqcondensate}  calculated with the solutions for $M(p^2)$ and $Z(p^2)$ using the
             longitudinal~\eqref{LVD} and transverse~\eqref{t-vertexstruc} vertex in the DSE~\eqref{finalDSEquark} and the three sets of gluon and ghost propagators
              introduced  in Section~\ref{ghostgluonsec}.}
              \label{table1}
\end{table}

\subsection{Applications}
\label{sec5}

While the numerical DSE solutions we obtain with the vertex defined by Eqs.~\eqref{LVD} and \eqref{t-vertexstruc} along with Eqs.~\eqref{lambda1QCD} to \eqref{tau8QCD}
lead to typical constituent masses and characteristic mass functions for all flavors considered in Figure~\ref{fig4}, only the calculation of a  gauge-independent
observable can tell us more about how realistic they are. Naturally, the next step consists in the construction of an antiquark-quark Bethe-Salpeter kernel that is consistent
with this new vertex and must satisfy the axialvector Ward-Takahashi identities. Given the complexity of the task, we postpone it to a future work.

However, even without the knowledge of the pion's Bethe-Salpeter amplitude we may compute its weak decay constant, which is a measure of chiral symmetry breaking.
Indeed, while the pion's mass vanishes in chiral limit, $f_\pi$ does not. We follow Ref.~\cite{Roberts:1994hh} where the weak decay constant in the chiral limit
is expressed by the integral,
\begin{align}
  \left  ( f_{\pi}^0 \right )^{\! 2} \,  = \,  \dfrac{N_c}{8 \pi^2}  \int_0^\infty \! dp^2 p^2  &\, B^2(p^2)  \left  (
                           \sigma_{\rm v}^2 - 2  \left [ \sigma_{\rm s}\sigma_{\rm s}^{\prime} +p^2 \sigma_{\rm v} \sigma_{\rm v}^{\prime} \right ]  \right.
 \nonumber \\
                     & \left.  - \ p^2 \left [ \sigma_{\rm s}\sigma_{\rm s}^{\prime \prime} - (\sigma_{\rm s}^{\prime} )^2 \right  ]
                                  - p^4 \left [ \sigma_{\rm v} \sigma_{\rm v}^{\prime \prime} - (\sigma_{\rm v}^{\prime} )^2 \right ]  \right ) \ ,
\label{fpi}
\end{align}
with $\sigma_{\rm s,v}^\prime \equiv d\sigma_{\rm s,v} (p^2)/dp^2$.
As we are constrained by a low renormalization point due to our use of quenched lattice-QCD input for the gluon and ghost propagators, determining  $Z_2$ and $Z_4$
in the chiral limit is not straightforward. We set $m= 0$~MeV at $\mu = 4.3$~GeV which allows for a sensible approximation in using $\sigma_{\rm s}(p^2)$ and 
$\sigma_{\rm v} (p^2)$ to calculate the decay constant in the chiral limit.

As another application, we consider the quark condensate which is an order parameter for DCSB. As for the pion decay constant, we do so using the three DSE solutions
for the full vertex at hand, i.e. $M(p^2)$ and $Z(p^2)$ obtained with the different ghost and gluon-dressing functions in the chiral limit. We thus calculate the integral over
the trace of the quark propagator:
\begin{equation}
   -  \langle\bar{q} q \rangle  \, \equiv \, Z_4  N_c  \int^{\Lambda}  \!  \frac{d^4k}{(2\pi)^4} \, \operatorname{tr}_{D}  \left [ S(k) \right ] \ .
 \label{qqcondensate}
\end{equation}
The values of the decay constant $f_\pi^0$ and of the quark condensate are obtained for all three sets of gluon and ghost propagators and are summarized
in Table~\ref{table1}. The three values for $f_\pi^0$ are slightly above the experimental value, $f_{\pi^\pm} = 92.2$~MeV, while the quark condensates are in good
agreement with other estimates, for instance the chiral condensate in the $\overline{\mathrm{MS}}$ scheme using SU(2) chiral perturbation theory~\cite{Borsanyi:2012zv},
$(-\langle \bar qq \rangle)^{1/3} =272(2)$~MeV, or the light-quark condensate from lattice QCD~\cite{McNeile:2012xh}: $(-\langle \bar qq \rangle)^{1/3} =283(2)$~MeV.


\section{Conclusive remarks and future developments}
\label{sec6}

We have derived a novel form of the transverse quark-gluon vertex that complements the ``longitudinal'' components obtained as a ghost-corrected Ball-Chiu vertex
in Refs.~\cite{Aguilar:2010cn,Aguilar:2016lbe,Aguilar:2018csq,Rojas:2013tza} which saturates the STI~\cite{Slavnov:1972fg,Taylor:1971ff}.
As for the non-transverse components, we were guided by symmetry transformations and multiplicative renormalizability encoded in the set of two TSTI~\cite{He:2009sj}
which couple the vector and axialvector vertices. By means of projections with two appropriate tensors one obtains an identity for each of
these vertices, i.e. the TSTI have been decoupled.

Once this is realized, one can project out the eight transverse form factors for which we obtain the expressions in Eqs.~\eqref{tau1QCD} to \eqref{tau8QCD}.
Notably, of the twelve form factors that describe the fully dressed quark-gluon vertex, only $\lambda_1 (k,p)$, $\lambda_2 (k,p)$, $\lambda_3 (k,p)$, $\tau_3 (k,p)$,
$\tau_5 (k,p)$ and  $\tau_8 (k,p)$ depend directly on the quark-ghost kernel and ghost-dressing function\footnote{\, $\lambda_4 (k,p)$ also depends on the
quark-ghost kernel if the form factors $X_1(k,p)$, $X_2(k,p)$ and $X_3(k,p)$ are not neglected as in the present case.}. Of course, due to imposing the
Bashir-Bermudez ansatz for the transverse vertex to constrain the scalar $Y_i(k,p)$ functions, the latter are also functions of the quark-ghost kernel
and ghost-dressing function. The last ingredients are the eight parameters, $\vec a :=\{a_1, a_2, a_3, a_4, a_5, a_6, a_7, a_8 \}$, inherent to the Bashir-Bermudez
vertex specified by Eqs.~\eqref{Rociotransverse1} to \eqref{Rociotransverse8}. We showed that they are far from being free parameters, as they are constrained by
multiplicative renormalizability relations and limited to a rather narrow range outside of which no or very unsatisfying solutions of the DSE are found.

Nonetheless, while our dressed quark-gluon vertex in its present form is successful in producing the right amount of DCSB for hadron phenomenology, as our
results for the weak decay constant of the pion and the quark condensate demonstrate, the reliance on parameters is unsatisfying and theoretically undesirable.
Amongst future perspectives, we plan to obtain an integral expression for the four-point function, at the origin of the tensor elements $T^{1}_{\mu\nu} V_{\mu\nu}$
and $T^{2}_{\mu\nu} V_{\mu\nu}$, along similar lines applied to the quark-ghost kernel in Section~\ref{quarkghost}. We remind that the nonlocal four-point function
is related to the tensor $V_{\mu\nu}$ via the line integral,
\begin{equation}
    V_{\mu\nu}  = \int\! \frac{d^4\ell}{(2\pi)^4}\, 2 \ell_\lambda\, \epsilon_{\lambda\mu\nu\rho}\,  \Gamma_\rho (k,p;\ell) \ ,
\label{lineintegral}
\end{equation}
where $ \Gamma_\rho (k,p;\ell) $ is the Fourier transform of the four-point function in coordinate space and which is defined in QED by~\cite{He:2009sj}:
\begin{align}
   \int \! d^4 x\, d^4 x' d^4 x_1\, d^4 x_2 & \ e^{i \left ( k \cdot x_1 - p \cdot x_2 + (p  -\ell) \cdot x- (k -\ell ) \cdot x' \right ) } \,
    \langle 0  |T \bar{\psi} (x' ) \gamma_{\rho} \mathcal{W} (x', x)  \psi(x) \psi (x_1) \bar{\psi} (x_2)  | 0 \rangle
\nonumber  \\
    = & \ (2 \pi)^4 \delta^4 (k-p-q ) \, S (k) \Gamma_\rho (k, p; \ell ) S (p) \ ,
\end{align}
with $q = (k -\ell) - (p- \ell)$ and $\mathcal{W} (x', x)$ is a Wilson line that ensures a gauge invariant  expression. The expression in QCD is analogous, yet involves
in addition color matrices and ghost fields. Therefore, expanding  the Wilson line to leading order in the strong coupling $g$~\cite{Costa:2021mpk},  the matrix element 
can be expressed approximately by diagrams that describe the gluon dressing of quarks, the two-quark scattering and the interaction of a quark with a 
ghost via gluon exchange. However, in such an approach the propagators  ought to be \emph{dressed\/} as in the dressed approximation to $H^a(k,p)$ discussed 
in Section~\ref{quarkghost}. With this ansatz, one can contract $V_{\mu\nu}$ with the tensors $T^{1}_{\mu\nu} $ and $T^{2}_{\mu\nu} $ to obtain the form factor 
decomposition of Eqs.~\eqref{TV1} and \eqref{TV2} and subsequently  project out all $Y_i(k,p)$ functions. As a consequence, we arrive at a set of eight integral 
equations which are all coupled with each other as well as with the integral equations for $A(p^2)$, $B(p^2)$ and $X_0(q^2)$ we considered in this work.

In future studies, the quark-ghost kernel we presented in Section~\ref{quarkghost} ought to be calculated beyond the leading approximation that includes only
$X_0(q^2)$ and moreover neglects any angular dependence; see Refs.~\cite{Aguilar:2016lbe,Aguilar:2018csq} for detailed calculations. This is important
 as $\lambda_4 (k,p)$ remains otherwise zero and in the particular case of the soft-gluon limit, $k=p$ ($q^2=0$), the expressions for
$\lambda_1(k,p)$~\eqref{lambda1QCD} and $\lambda_2(k,p)$~\eqref{lambda2QCD} hardly express any functional dependence without the full inclusion of the
$X_i(k,p)$~\cite{Oliveira:2020yac}. However, for our present purpose this simplified ansatz proves to be sufficient, as our aim was to demonstrate that an important
amount of DCSB in the gap equation is due to the transverse quark-gluon vertex. We are therefore optimistic that with future refinements a complete vertex structure
can be achieved that catches the dynamical subtleties much beyond the leading truncation, indispensable for the reproduction of the excited hadron spectrum
and exotic states.


\acknowledgments

We kindly thank Orlando Oliveira for helpful comments about the refined Gribov-Zwanziger parametrization of the ghost- and gluon-dressing functions
and for providing the corresponding lattice QCD data. B.E. acknowledges funding by FAPESP, grant no.~2018/20218-4, and by CNPq,
grant no.~428003/2018-4. F.E.S. is supported by CAPES-PNPD grant no.~88882.314890/2013-01, R.C.S. by a CAPES PhD fellowship and L.A. by a 
FAPESP postdoctoral fellowship grant no.~2018/17643-5.
E.R. acknowledges support from ``Vicerrector\'ia de Investigaciones e Interacci\'on Social VIIS de la Universidad de Nari\~no'', project numbers 1928 and 2172.
This research was also partly supported by Coordinaci\'on de la Investigaci\'on Cient\'ifica (CIC) of the University of Michoacan and CONACyT, Mexico, through 
grant nos.~4.10 and CB2014-22117, respectively. This work is part of the project ``INCT-F\'isica Nuclear e Aplica\c{c}\~oes'', no. 464898/2014-5.


\end{document}